\documentclass[conference]{IEEEtran}

\makeatletter
\newcommand{\linebreakand}{%
  \end{@IEEEauthorhalign}
  \hfill\mbox{}\par
  \mbox{}\hfill\begin{@IEEEauthorhalign}
}
\makeatother

\usepackage{utils}
\usepackage{tikz}
\usepackage{paralist}
\usepackage{tcolorbox}

\hypersetup{
    colorlinks,%
    citecolor=black,%
    filecolor=black,%
    linkcolor=black,%
    urlcolor=black}

\newtcolorbox{resultbox}{colback=lightgray, arc=0.5mm, top=1mm, bottom=1mm, left=1mm, right=1mm}

\begin{document}

%
% paper title
% Titles are generally capitalized except for words such as a, an, and, as,
% at, but, by, for, in, nor, of, on, or, the, to and up, which are usually
% not capitalized unless they are the first or last word of the title.
% Linebreaks \\ can be used within to get better formatting as desired.
% Do not put math or special symbols in the title.

\title{\emph{``Let it be Chaos in the Plumbing!"} Usage and Efficacy of Chaos Engineering in DevOps Pipelines}

\author{\IEEEauthorblockN{Stefano Fossati}
\IEEEauthorblockA{\textit{Jheronimus Academy of Data Science} \\
\textit{Eindhoven University of Technology} \\
Eindhoven 5612AZ, Netherlands \\
s.fossati@tue.nl}
\and
\IEEEauthorblockN{Damian Andrew Tamburri}
\IEEEauthorblockA{\textit{Department of Engineering} \\
\textit{University of Sannio \& JADS/NXP Semiconductors} \\
Benevento 82100, Italy \\
datamburri@unisannio.it}
\linebreakand
\IEEEauthorblockN{Massimiliano Di Penta}
\IEEEauthorblockA{\textit{Department of Engineering} \\
\textit{University of Sannio} \\
Benevento 82100, Italy \\
dipenta@unisannio.it}
\and
\IEEEauthorblockN{Marco Tonnarelli}
\IEEEauthorblockA{\textit{Jheronimus Academy of Data Science} \\
\textit{Eindhoven University of Technology} \\
Eindhoven 5612AZ, Netherlands \\
m.tonnarelli@tue.nl}
}

\maketitle
\thispagestyle{empty}

\begin{abstract}
Chaos Engineering (CE) has emerged as a proactive method to improve the resilience of modern distributed systems, particularly within DevOps environments. Originally pioneered by Netflix, CE simulates real-world failures to expose weaknesses before they impact production. In this paper, we present a systematic gray literature review that investigates how industry practitioners have adopted and adapted CE principles over recent years. Analyzing 50 sources published between 2019 and early 2024, we developed a comprehensive classification framework that extends the foundational CE principles into ten distinct concepts. Our study reveals that while the core tenets of CE remain influential, practitioners increasingly emphasize controlled experimentation, automation, and risk mitigation strategies to align with the demands of agile and continuously evolving DevOps pipelines. Our results enhance the understanding of how CE is intended and implemented in practice, and offer guidance for future research and industrial applications aimed at improving system robustness in dynamic production environments.
\end{abstract}
\IEEEpeerreviewmaketitle

\section{Introduction}
Chaos Engineering (CE) has emerged as a discipline aimed at simulating failures with the aim of understanding the system's capability to tolerate them before they manifest in production environments~\cite{basiri_chaos_2016}, typically within the context of DevOps pipelines and connected tools. 
Basiri and the Netflix Chaos team~\cite{basiri_chaos_2016} defined CE and its principles as 
\begin{quote}
    \textit{Modern software-based services are implemented as distributed systems with complex behavior and failure modes. "Chaos engineering uses experimentation to ensure system availability." Netflix engineers have developed principles of chaos engineering that describe how to design and run experiments.}
\end{quote}

CE has garnered substantial interest in both industry~\cite{gremlin_state_2021} and academia in recent years~\cite{rosenthal_chaos_2020}. The reasons are manifold. On the one hand, modern information systems are becoming increasingly complex~\cite{sommerville_large-scale_2012}. For example, according to Deloitte and Mulesoft Report (2024)~\cite{mulesoft_2024_nodate}, the average enterprise IT environment in the retail and consumer goods sector uses 1,022 different applications to manage operations, creating highly distributed architectures with numerous interconnections. On the other hand, modern systems' complexity introduces significant DevOps operational challenges. For example, outages are often caused by unexpected interactions between components in these highly distributed environments, leading to severe production disruptions that could negatively impact the users. In 2024, 25\% of respondents worldwide reported that the average hourly downtime cost of their servers ranged between \$301,000 and \$400,000~\cite{statista_global_nodate}. While academic studies have focused on various aspects of CE for domain-specific software maintenance~\cite{9831186} and continued operational capacity~\cite{10428037}, its application in industrial DevOps practice has been steadily advancing. The growing use of CE in the corporate domain is reflected in gray literature, such as blogs and articles written by practitioners. However, we identified a gap in the literature concerning a systematic understanding of how practitioners perceive and prioritize key aspects of CE in action. To address this gap, this paper aims to explore the aspects practitioners consider most significant in their industrial practice, discussing the main CE concepts and binding them with DevOps concepts and practice. We systematically analyzed 50 gray literature sources published between 2019 and April 2024, following the guidelines of Garousi~\cite{garousi_guidelines_2019}. This analysis led to the creation of a classification framework highlighting the aspects practitioners deem important in CE. This study contributes to both academia and industry by providing an overview of practitioner perspectives on the main elements of CE. Additionally, it offers practical insights into how CE can interact with and complement DevOps practices.

The paper is structured as follows. Section~\ref{sec:ce_state} reports related work of CE. Section~\ref{sec:research_methodology} illustrates the methodology, research questions, and scope of our systematic study. Section~\ref{sec:glr_results} presents the results of our study, while Section~\ref{sec:glr_discussion} reports the outcome of the validation of our findings with a software development company. Threats to the study's validity are discussed in Section~\ref{sec:threats_of_validity}, while Section~\ref{sec:glr_conclusions} concludes the paper and outlines its directions for future work.

\section{Chaos Engineering: a Digest}
\label{sec:ce_state}

The foundational principles outlined in the Netflix article~\cite{basiri_chaos_2016} served as a basis for subsequent academic research on CE. 

The scholars' community began exploring various perspectives on chaos engineering, including its business implications.  Tucker et al.~\cite{tucker_business_2018} analyzed the cost-related benefits of CE, quantitatively assessing its impact on organizational performance. By analyzing historical data, they identified instances where CE could yield tangible benefits to businesses.

Since chaos engineering requires observability of the system, several studies have focused on enhancing system observability. Simonsson et al.~\cite{simonsson_observability_2021} developed a tool to provide observability for analyzing the results of chaos experiments on Docker, while Zhang et al.~\cite{zhang_automatic_2021} examined observability requirements for Docker Java applications, particularly in resilience and chaos engineering contexts, proposing an approach tailored to such applications. 

Based on the observability of the system call invocations, Zhang et al~\cite{zhang_maximizing_2022} develop a tool to maximize the realism of the chaos injections based on the errors that naturally happen in the production environment. 

On the side of the type of chaos experiments that should be done on a system, Kesim et al.~\cite{kesim_identifying_2020} propose a systematic approach to chaos engineering to prioritize the experiments based on risk analysis techniques on the system. More focused on the experiments with network fault injection, Cotroneo et al.~\cite{cotroneo_thorfi_2022} developed a non-intrusive fault injection tool (ThorFI) for virtual networks in a cloud computing infrastructure. This tool is based on their design solution for cloud tenants to support them in assessing the resiliency of their cloud applications, while avoiding side effects on the other tenants. Ikeuchi et al.~\cite{ikeuchi_coverage_2023} focused on improving the efficiency of chaos engineering and proposed an algorithm to give exact optimal solutions to coverage maximization and iteration minimization problems when imposing connectivity as an invariant.

One of the chaos engineering principles is focused on doing chaos experiments in a production environment. So some researchers have explored methods for simulating chaos experiments in controlled environments to avoid disruptions in production. Frank et al.~\cite{frank_misim_2022} introduced a simulator for microservice architectures, allowing users to conduct chaos experiments in a controlled and safe environment. Additionally, some researchers have proposed using the \textit{digital twins} concept based on the creation of a digitally accurate counterpart of the system. They think that it  ``would bring many of the advantages of the Chaos Engineering approach with a much smaller price tag and enabling much faster feedback cycles at the design level'' ~\cite{poltronieri_chaostwin_2021}  ~\cite{fogli_chaos_2024}.

Other researchers focused their attention on the automation stuff according to the last principle mentioned in the section before. Alvaro et al.~\cite{alvaro_automating_2016} and successively Basiri et al.~\cite{basiri_automating_2019} described how they have automated chaos experiments in Netflix, which manages a very complex distributed system.

Jernberg et al.~\cite{jernberg_getting_2020} developed a framework for guiding the implementation of chaos engineering inside a company. In particular, they conceptualized the problem by considering the literature studies of chaos engineering and interviewing the practitioners of the case study company. In a second moment, they validated the solution framework by implementing chaos engineering inside a specific company.

\section{Methodology}
\label{sec:research_methodology}
In this section, we describe the study method, which is based on gray literature analysis.
The inclusion of gray material in systematic reviews has become popular in software engineering~\cite{garousi_benefitting_2020} due to the value that this online information can bring in academic research~\cite{farace_grey_2010,garousi_need_2016}. Following the guidelines provided by Kitchenham et al.~\cite{kitchenham_guidelines_2007} and Garousi et al.~\cite{garousi_guidelines_2019}, we conducted systematic reviews involving gray literature. 
We adopt these guidelines and integrate them with the approach outlined for conducting systematic literature reviews in software engineering by Petersen et al.~\cite{kai_petersen_guidelines_2015}. Online resources are important for bringing direct insights and experiences from software engineering practitioners and capturing diverse perspectives. 

\begin{figure}
    \centering
    \includegraphics[width=1\linewidth]{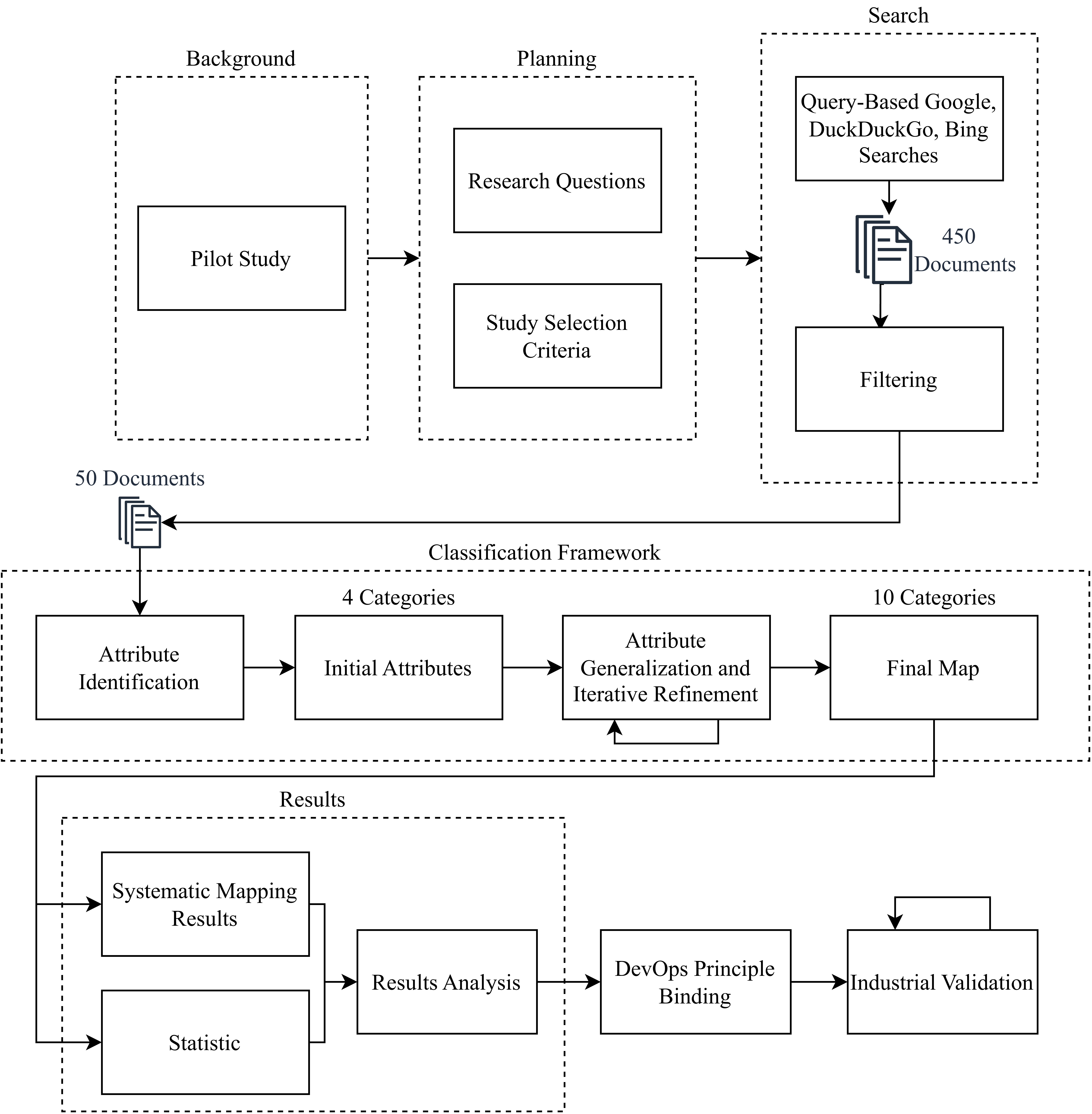}
    \caption{Methodology, an overview.}
    \label{fig:FrameworkDiag}
    \vspace{-0.3mm}
\end{figure}

As shown in Fig.~\ref{fig:FrameworkDiag},  the process we followed is structured as follows, according to the guidelines mentioned above:
\begin{compactitem}
    \item The search process is done on research engines, and the search stops with the saturation criteria;
    \item The sources found with the research process are selected following inclusion and exclusion criteria;
    \item The data is extracted following a classification framework; and
    \item The results obtained from the systematic mapping and statistics are analyzed and discussed.
\end{compactitem}

Finally, to elicit practical considerations and observations reflecting the relation between our findings and their real-life exploitation in an industrial context, we validated results and findings in two industrial workshops (see Section~\ref{sec:glr_discussion}).

\subsection{Research problem definition and research questions}
\label{subsec:research_problem_definition}

The \emph{goal} of our study is to evaluate experienced practitioners' perspectives on the CE concepts and the importance they attribute to them. Overall, we address the following research question:
\textbf{RQ: \emph{``What generalizable concepts drive Chaos Engineering?"}}

\subsection{Research queries and selection criteria}
\label{research_queries}

Starting from the research question, we define the research queries that allow us to find the gray material on the web. As suggested by Petersen et al.~\cite{kai_petersen_systematic_2008}, we identified the search string using our research question RQ. More precisely, we focused on the Population and Intervention terms of the PICO method defined by Kitchenham et al.~\cite{kitchenham_guidelines_2007}. Our query~(\ref{query}) is designed to search for online materials authored by practitioners and experts discussing CE concepts and their importance, and the keywords are extracted by RQ.

\begin{equation}
\label{query}
    \begin{gathered}
        (\textrm{\textit{chaos}} \wedge \textrm{\textit{engineering}}) \wedge
        (\textrm{\textit{principle*}} \vee \textrm{\textit{best practice*}} \vee \\\textrm{\textit{guideline*}} \vee
        \textrm{\textit{key concept*}} )
    \end{gathered}
\end{equation}

We conducted our search primarily on mainstream search engines, including Google, DuckDuckGo, and Bing. For each engine query---as instantiated over the specific search engine with its own specialized query syntax---we collected the first 30 results for each subqueries, resulting in a total of 450 results. Note that for each search engine, we opted to select the first 30 (i.e., most relevant, determined by focusing on the top 30 results of our subqueries above) results to balance the completeness of our findings with the manageability of the data volume.

Then, we excluded duplicates using a script, leaving us with a total of 181 unique results. During the deduplication phase, we also manually verified that different resources had different authors to prevent redundancy in content authored by the same individual. Additionally, we ensured that resources hosted by the same domain were written by different authors and did not contain identical content. To ensure information quality, we filtered the gray sources based on criteria outlined in Table \ref{table:SelectionCriteria}. Sources were included if they met any of the inclusion criteria and excluded if they failed to meet at least one exclusion criterion.
\begin{table}
\caption{Inclusion and Exclusion Criteria.}
\label{table:SelectionCriteria}
    \centering
    \small
    \begin{tabular}{p{0.07\textwidth}|m{0.01\textwidth}m{0.33\textwidth}}
    \hline
%    \rowcolor{bluepoli!40}
    \textbf{Criteria} &  & \textbf{Criteria Description} \\
    \hline 
    \multirow{4}{*}{Inclusion} & I1 &  The source is released by company that uses Chaos Engineering  \\ \cline{2-3}
    & I2 & The source is released by a company that sells Chaos Engineering tools  \\ \cline{2-3}
    & I3 & The source is written by authors with quantitative experience in software engineering \\ \cline{2-3}
    & I4 & The source is considered fundamental by the Chaos Engineering community \\ 
    \hline
    \multirow{5}{*}{Exclusion} & E1 & The source is written before 2016 \\ \cline{2-3}
    & E2 & The source is not written in English \\ \cline{2-3}
    & E3 & The source is not text or is not part of the gray sources \\ \cline{2-3}
    & E4 & The source is not free accessible \\ \cline{2-3}
    & E5 & The source is not related to Chaos Engineering principles and concepts\\
    \hline
    \end{tabular}
    \vspace{-0.1mm}
\end{table}
We opted to include sources from companies utilizing CE as they offer valuable insights into enterprise experiences with this practice. 

All sources authored by individuals with experience in software engineering were included to extract insights from practitioners. We assessed the author's experience using LinkedIn, with a particular focus on technical individuals who have years of experience or are currently active in IT environments.
Furthermore, we included online sources considered fundamental within the CE community, specifically the articles by (some of) the original authors of the Chaos Engineering paper of Netflix or by a contributor who posted to the Chaos Engineering website\footnote{\url{https://principlesofchaos.org/}}. This is because such a website indexes all contributions deemed relevant by the community to chaos engineering. Considering the exclusion criteria, we initially excluded sources written before 2016, as this marks the year when CE was first defined by Basiri et al.~\cite{basiri_chaos_2016}. The publication date of the blog post was retrieved from the website or its header.  We excluded sources not in English and those that were not fully accessible. Non-text resources and peer-reviewed materials, such as papers and books on CE, were also excluded. Finally, any sources not discussing CE principles or related topics were excluded as they do not contribute to addressing our research questions. To ensure transparency and reproducibility, we provide a replication package\footnote{https://doi.org/10.5281/zenodo.15846570}.

\begin{table*}[t!]
    \caption{Mapping of the source type and the resource selected.}
    \label{tab:source_type}
    \small
    \begin{tabular}{m{0.10\textwidth}|m{0.05\textwidth}|>{\arraybackslash}m{0.75\textwidth}}
    \hline
    Source Type & Count & References
    \\ \hline
    Ad hoc blog & 1 &~\cite{shelton_what_2022}
    \\ \hline
    Community & 17   &~\cite{singh_art_2023},  ~\cite{arvind_chaos_2021},~\cite{rosenthal_what_2021},  ~\cite{chibuike_chaos_2023},~\cite{hornsby_what_2020},  ~\cite{preet_what_2024},~\cite{chattopadhyay_guide_2024},  ~\cite{hanmer_chaos_2024},~\cite{green_7_2023},  ~\cite{kalal_day50-_2024},~\cite{sanwal_chaos_2023},  ~\cite{treat_guidelines_2020},~\cite{ukkuru_chaos_2023},  ~\cite{keemick_quick_2023},~\cite{nino_roa_chaos_2022},  ~\cite{tricentis_chaos_2022},~\cite{community_principles_2019} 
    \\ \hline
    Vendor  & 32    &~\cite{ruqayya_chaos_2022},  ~\cite{varma_mastering_2024},~\cite{aws_rel12-bp05_nodate},  ~\cite{gremlin_chaos_2023},  ~\cite{claytonsiemens77_recommendations_2023},  ~\cite{kareliya_what_2023},~\cite{prasla_embracing_2024},  ~\cite{maheshwari_chaos_2023},~\cite{clifford_unleash_2022},  ~\cite{andrades_what_2023},~\cite{lella_practical_2022},  ~\cite{dantoni_what_2022},~\cite{abdul_vault_2024},  ~\cite{wakayama_article_2020}~\cite{schillerstrom_what_2022}  ~\cite{bairyev_unlocking_2023}~\cite{j_haber_why_2021}  ~\cite{parekh_building_2021},~\cite{prithvi_chaos_2021},~\cite{bremmers_how_2021},~\cite{gianchandani_chaos_2022},  ~\cite{kadikar_building_2023},~\cite{ahamed_chaos_2023},  ~\cite{bhaskar_chaos_2022},~\cite{eliot_verify_2022}, ~\cite{gunja_what_2024},~\cite{wickramasinghe_chaos_2023},~\cite{singh_gill_chaos_2021},~\cite{apexon_what_2020},  ~\cite{adservio_chaos_2021},~\cite{opentext_what_nodate},  ~\cite{ibm_what_2024} \\ \hline
    \end{tabular}
    \vspace{-0.1mm}
\end{table*}

\subsection{Data Extraction}
\label{subsec:data_extraction}
To conduct a systematic mapping, we followed the approach by Garousi et al.~\cite{garousi_guidelines_2019} and, consequently,   the guidelines of Kai Petersen et al.~\cite{kai_petersen_systematic_2008}. Our classification scheme started with an initial version and evolved during data extraction through attribute generalization and iterative refinement steps. New categories were added, and existing categories were merged or split. Once the iterative refinement process was finished and the scheme consistently categorized all the sources in the pool, a stable final scheme was derived. During the classification of the resources, we categorized them by year of publication (where available, either in the resource itself or in its metadata), and type of source, i.e.:

\begin{compactenum}
    \item \textit{Communities}, i.e.,  platforms such as LinkedIn\footnote{https://www.linkedin.com/}, Medium\footnote{https://medium.com/} Devopedia\footnote{https://devopedia.org/}, and others.
    \item \textit{Ad hoc blogs} Where specific individuals or organizations share their own knowledge.
    \item \textit{Vendor websites}: Where companies sell products, featuring sections where relevant individuals can share knowledge.
\end{compactenum}

We focused on the principles from which we started the classification. Such principles, defined by Basiri et al.~\cite{basiri_chaos_2016}, are:
\begin{compactenum}
\item \textbf{P1. Define The Steady-State Hypothesis:}  Identify and measure Key Performance Indicators (KPIs) to define the system's normal state.
\item \textbf{P2. Vary real-world events:} – Simulate realistic failures (e.g., service crashes, network latency) to test system resilience.
\item \textbf{P3. Run Experiments in Production:} – Test directly in the production environment for reliable results.
\item \textbf{P4. Automate Experiments to Run Continuously:} – Ensure continuous testing to adapt to system updates automatically.
\end{compactenum}

Throughout the classification step,  we manually conducted a thematic analysis to extract relevant concepts from each resource. We applied manual coding to analyze the raw material, iteratively assigning each piece of information to predefined categories based on the CE principles (P1–P4). When a concept did not fit existing categories, we created new ones, refining our taxonomy in successive iterations. And when a principle encompassed multiple distinct aspects, we split it into more detailed subcategories.

\section{Study Results}
\label{sec:glr_results}

\begin{figure}
    \centering
    \includegraphics[width=0.9\linewidth]{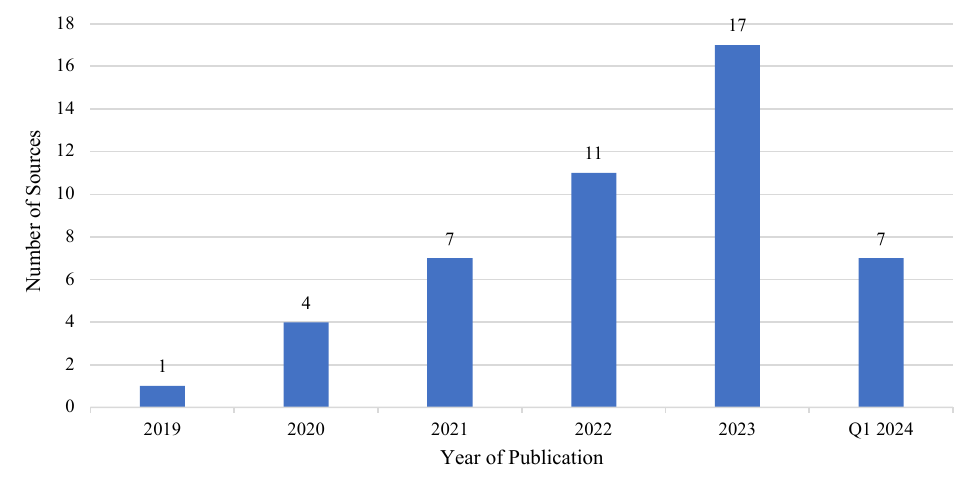}
    \caption{Distribution of the selected gray sources per year from 2019 to Q1 2024.}
    \label{fig:SourceYearDist}
    \vspace{-0.1mm}
\end{figure}

\begin{table}[t!]
    \centering
    \small
    \caption{Mapping of Chaos Engineering Concepts of Practitioner and DevOps Concepts.}
    \label{tab:mapping}
    \begin{tabular}{|c|c|c|}
    \hline
    \textbf{CE Principles~\cite{basiri_chaos_2016}} &
    \textbf{CE Concepts} & \textbf{DevOps Concepts} \\ \hline
    P1 & C1 & Runtime \\ \hline
    P1 & C2 & Process \\ \hline
    P2 & C3 & Runtime \\ \hline
    P3 & C4 & Runtime \\ \hline
    P4 & C5 & Delivery \\ \hline
    - & C6 & Runtime/People \\ \hline
    - & C7 & Process/Runtime \\ \hline
    - & C8 & Process \\ \hline
    - & C9 & People \\ \hline
    - & C10 & Process \\ \hline
    \end{tabular}
    \vspace{-0.1mm}
\end{table}

\begin{figure}
    \centering
    \includegraphics[width=0.85\linewidth]{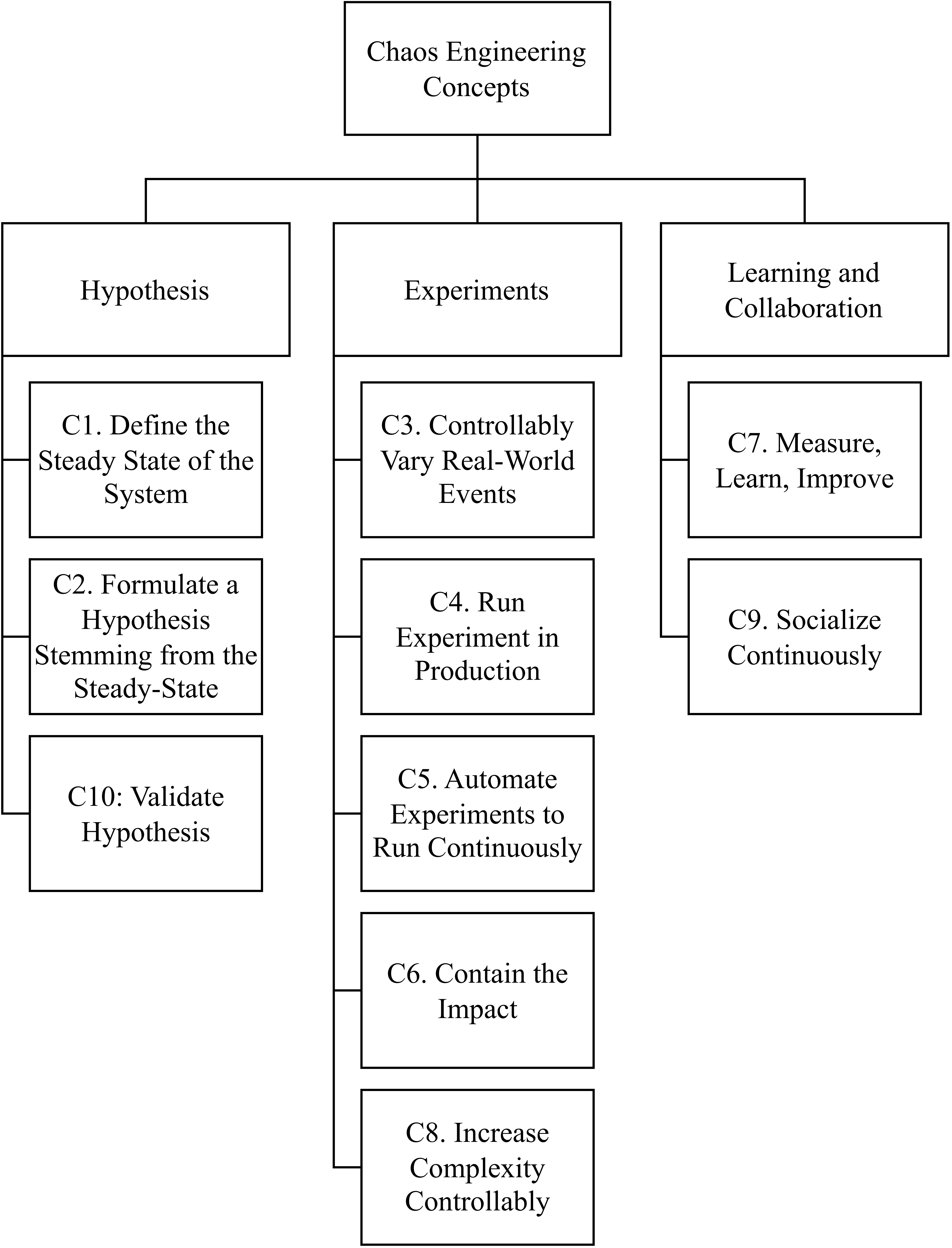}
    \caption{Taxonomy of relevant Chaos Engineering concepts elicited from gray literature.}
    \label{fig:Taxonomy}
    \vspace{-0.1mm}
\end{figure}

\subsection{Results overview}
\label{subsec:glr_results_overview}
The search engine has retrieved sources that start from 2019 to 2024 (in particular, May 2024, which is the month in which the search was conducted and, therefore, where the analysis stops). As shown in Fig.~\ref{fig:SourceYearDist}, the online resources referring to CE have been growing in the last few years. Table \ref{tab:source_type} reports the references to all considered resources along different source types. As shown, the majority are blog posts from tool vendors (32 of them), whereas 17 come from community blogs. Only one resource is an ad-hoc blog. 

\subsection{Main Chaos Engineering Concepts for Practitioners}
\label{subsec:practical_principles}
In this section, we report and discuss the classification of the analyzed sources along the four principles
defined by Garousi et al.~\cite{garousi_need_2016} as shown in Table~\ref{tab:mapping}. This Table also shows a mapping between the DevOps and CE extracted concepts that will be discussed in Section~\ref{subsec:mapping}
This led, as shown in Fig.~\ref{fig:Taxonomy}, 
to a taxonomy of  10 concepts, in turn organized into three high-level groups.
In the following, we describe the taxonomy concepts along the three high-level categories. Note that concepts do not follow a subsequent order along categories, as we kept the ordering from our original coding.

\paragraph{\textbf{Hypothesis}} This category focuses on the formulation of hypotheses to guide CE experiments. It involves identifying the steady state of the system and defining expectations about its behavior under adverse conditions. By answering ``What can go wrong?'' practitioners establish clear criteria for failure and success, supported by measurable Key Performance Indicators (KPIs). The hypotheses provide a scientific framework for experimentation and analysis. This category features three concepts:

\begin{compactenum}
 \item \textbf{C1. Define the Steady State of the System.} This concept identifies the normal behavior of the target system that is registered when the system is working in normal conditions (e.g., no network, no overload, everything is working as expected).  
    As  Basiri et al.~\cite{basiri_chaos_2016} reported, the "normal" behavior should be defined with specific KPI that are strictly related to the specific system. In an application that is used for streaming, a good KPI can be Stream starts per second, or in an application that has to manage a database (DB), writing it can be Write per second. In the paper by Basiri et al.~\cite{basiri_chaos_2016}, this concept is attached to the next one that we define. However, to perform  a more fine-grained classification, we preferred to split them:
    \item \textbf{C2. Formulate a Hypothesis Stemming from the Steady-State.} In this stage, the practitioner should build a hypothesis in case of some problems connected with the target system. The hypothesis typically answers the question: ``What can go wrong?'' within the target system. This hypothesis should be detailed with some KPI expected from the failure. Basiri et al.~\cite{basiri_chaos_2016} specify that the KPI that should be taken into consideration should be connected to the effect on the user side, not to the backend side of the application. Netflix reasoned about what they want to avoid, so as to affect the user's experience of the service. So their KPI are connected to some parameter that directly represents the effect on the user experience (e.g., for a video streaming service like Netflix, they have considered as KPI the frame per second as a KPI to measure the quality of service to users.)
  \item \textbf{C10. Validate Hypothesis.} This concept has more of a practical than a theoretical aspect. It is more related to the execution of the experiment and the validation of the hypothesis after the experiment is concluded, but we can separate it to have a more refined classification. This concept was also reported by Basiri et al.~\cite{basiri_chaos_2016} as a step for the experiment execution.
\end{compactenum}

\begin{table*}[t]
    \caption{Association of the concepts found in the selected resources}
    \label{tab:concepts_sources}
    
    \scriptsize
    \centering
    \begin{tabular}{m{0.03\textwidth}|>{\arraybackslash}m{0.9\textwidth}}
    \hline
    \textbf{Cncpt} & \textbf{Sources}      
    \\ \hline
    C1      &~\cite{ruqayya_chaos_2022},  ~\cite{varma_mastering_2024},  ~\cite{aws_rel12-bp05_nodate},  ~\cite{claytonsiemens77_recommendations_2023},  ~\cite{kareliya_what_2023},~\cite{prasla_embracing_2024},  ~\cite{singh_art_2023},~\cite{arvind_chaos_2021},  ~\cite{maheshwari_chaos_2023},~\cite{rosenthal_what_2021},  ~\cite{clifford_unleash_2022},~\cite{andrades_what_2023},  ~\cite{hornsby_what_2020},~\cite{lella_practical_2022},  ~\cite{dantoni_what_2022},~\cite{abdul_vault_2024},  ~\cite{preet_what_2024},~\cite{wakayama_article_2020},  ~\cite{schillerstrom_what_2022},  ~\cite{bairyev_unlocking_2023},~\cite{j_haber_why_2021},  ~\cite{parekh_building_2021},~\cite{gianchandani_chaos_2022},  ~\cite{kadikar_building_2023},~\cite{ahamed_chaos_2023},  ~\cite{bhaskar_chaos_2022},~\cite{eliot_verify_2022},  ~\cite{gunja_what_2024},~\cite{wickramasinghe_chaos_2023},  ~\cite{chattopadhyay_guide_2024},~\cite{hanmer_chaos_2024},  ~\cite{green_7_2023},~\cite{kalal_day50-_2024},  ~\cite{sanwal_chaos_2023},~\cite{treat_guidelines_2020},  ~\cite{nino_roa_chaos_2022},~\cite{singh_gill_chaos_2021},  ~\cite{apexon_what_2020},~\cite{tricentis_chaos_2022},  ~\cite{adservio_chaos_2021},~\cite{community_principles_2019},  ~\cite{ibm_what_2024}
    \\ \hline
    C2      &~\cite{ruqayya_chaos_2022},~\cite{varma_mastering_2024},  ~\cite{aws_rel12-bp05_nodate},  ~\cite{gremlin_chaos_2023},  ~\cite{claytonsiemens77_recommendations_2023},  ~\cite{kareliya_what_2023},~\cite{prasla_embracing_2024},  ~\cite{singh_art_2023},~\cite{arvind_chaos_2021},  ~\cite{maheshwari_chaos_2023},~\cite{rosenthal_what_2021},  ~\cite{chibuike_chaos_2023},~\cite{clifford_unleash_2022},  ~\cite{andrades_what_2023},~\cite{hornsby_what_2020},  ~\cite{lella_practical_2022},~\cite{dantoni_what_2022},  ~\cite{abdul_vault_2024},~\cite{preet_what_2024},  ~\cite{wakayama_article_2020},~\cite{shelton_what_2022},  ~\cite{schillerstrom_what_2022},  ~\cite{bairyev_unlocking_2023},~\cite{j_haber_why_2021},  ~\cite{parekh_building_2021},~\cite{prithvi_chaos_2021},  ~\cite{bremmers_how_2021},~\cite{gianchandani_chaos_2022},  ~\cite{kadikar_building_2023},~\cite{ahamed_chaos_2023},  ~\cite{bhaskar_chaos_2022},~\cite{eliot_verify_2022},  ~\cite{gunja_what_2024},~\cite{wickramasinghe_chaos_2023},  ~\cite{chattopadhyay_guide_2024},~\cite{hanmer_chaos_2024},  ~\cite{green_7_2023},~\cite{kalal_day50-_2024},  ~\cite{sanwal_chaos_2023},~\cite{treat_guidelines_2020},  ~\cite{keemick_quick_2023},~\cite{nino_roa_chaos_2022},~\cite{singh_gill_chaos_2021},  ~\cite{apexon_what_2020},~\cite{tricentis_chaos_2022},  ~\cite{community_principles_2019},~\cite{opentext_what_nodate} 
    \\ \hline
    C3      &~\cite{ruqayya_chaos_2022},  ~\cite{aws_rel12-bp05_nodate},  ~\cite{claytonsiemens77_recommendations_2023},  ~\cite{kareliya_what_2023},~\cite{prasla_embracing_2024},  ~\cite{singh_art_2023},~\cite{arvind_chaos_2021},  ~\cite{maheshwari_chaos_2023},~\cite{rosenthal_what_2021},  ~\cite{chibuike_chaos_2023},~\cite{clifford_unleash_2022},  ~\cite{andrades_what_2023},~\cite{hornsby_what_2020},  ~\cite{preet_what_2024},~\cite{wakayama_article_2020},  ~\cite{shelton_what_2022},~\cite{schillerstrom_what_2022},  ~\cite{bairyev_unlocking_2023},~\cite{j_haber_why_2021},  ~\cite{parekh_building_2021},~\cite{prithvi_chaos_2021},  ~\cite{ahamed_chaos_2023},~\cite{bhaskar_chaos_2022},  ~\cite{eliot_verify_2022},~\cite{wickramasinghe_chaos_2023},  ~\cite{sanwal_chaos_2023},~\cite{ukkuru_chaos_2023},  ~\cite{nino_roa_chaos_2022},~\cite{apexon_what_2020},  ~\cite{tricentis_chaos_2022},~\cite{adservio_chaos_2021},  ~\cite{community_principles_2019},~\cite{ibm_what_2024}
    \\ \hline
    C4      &~\cite{claytonsiemens77_recommendations_2023},  ~\cite{arvind_chaos_2021},~\cite{maheshwari_chaos_2023},  ~\cite{clifford_unleash_2022},~\cite{wakayama_article_2020},  ~\cite{shelton_what_2022},~\cite{schillerstrom_what_2022},  ~\cite{j_haber_why_2021},~\cite{ahamed_chaos_2023},  ~\cite{wickramasinghe_chaos_2023},~\cite{hanmer_chaos_2024},  ~\cite{singh_gill_chaos_2021},~\cite{tricentis_chaos_2022},  ~\cite{adservio_chaos_2021},~\cite{community_principles_2019},  ~\cite{ibm_what_2024}
    \\ \hline
    C5      &~\cite{ruqayya_chaos_2022},~\cite{varma_mastering_2024},  ~\cite{aws_rel12-bp05_nodate},~\cite{singh_art_2023},  ~\cite{arvind_chaos_2021},~\cite{maheshwari_chaos_2023},  ~\cite{clifford_unleash_2022},~\cite{dantoni_what_2022},  ~\cite{wakayama_article_2020},~\cite{schillerstrom_what_2022},  ~\cite{bairyev_unlocking_2023},~\cite{j_haber_why_2021},  ~\cite{prithvi_chaos_2021},~\cite{bremmers_how_2021},  ~\cite{bhaskar_chaos_2022},~\cite{eliot_verify_2022},  ~\cite{wickramasinghe_chaos_2023},  ~\cite{chattopadhyay_guide_2024},~\cite{green_7_2023},  ~\cite{kalal_day50-_2024},~\cite{ukkuru_chaos_2023},~\cite{nino_roa_chaos_2022},  ~\cite{singh_gill_chaos_2021},~\cite{tricentis_chaos_2022},  ~\cite{community_principles_2019},~\cite{ibm_what_2024} 
    \\ \hline
    C6      &~\cite{ruqayya_chaos_2022},~\cite{varma_mastering_2024},  ~\cite{aws_rel12-bp05_nodate},  ~\cite{gremlin_chaos_2023},  ~\cite{claytonsiemens77_recommendations_2023},  ~\cite{singh_art_2023},~\cite{arvind_chaos_2021},  ~\cite{maheshwari_chaos_2023},~\cite{chibuike_chaos_2023},  ~\cite{clifford_unleash_2022},~\cite{andrades_what_2023},  ~\cite{hornsby_what_2020},~\cite{lella_practical_2022},  ~\cite{abdul_vault_2024},~\cite{wakayama_article_2020},  ~\cite{schillerstrom_what_2022},  ~\cite{bairyev_unlocking_2023},~\cite{j_haber_why_2021},  ~\cite{prithvi_chaos_2021},~\cite{gianchandani_chaos_2022},  ~\cite{bhaskar_chaos_2022},~\cite{eliot_verify_2022},  ~\cite{gunja_what_2024},~\cite{wickramasinghe_chaos_2023},  ~\cite{chattopadhyay_guide_2024},~\cite{hanmer_chaos_2024},  ~\cite{green_7_2023},~\cite{kalal_day50-_2024},  ~\cite{keemick_quick_2023},~\cite{singh_gill_chaos_2021},  ~\cite{apexon_what_2020},~\cite{tricentis_chaos_2022},  ~\cite{adservio_chaos_2021},~\cite{community_principles_2019},  ~\cite{ibm_what_2024}
    \\ \hline
    C7      &~\cite{ruqayya_chaos_2022},~\cite{varma_mastering_2024},  ~\cite{aws_rel12-bp05_nodate},  ~\cite{gremlin_chaos_2023},  ~\cite{claytonsiemens77_recommendations_2023},  ~\cite{prasla_embracing_2024},~\cite{singh_art_2023},  ~\cite{maheshwari_chaos_2023},~\cite{chibuike_chaos_2023},  ~\cite{clifford_unleash_2022},~\cite{andrades_what_2023},  ~\cite{hornsby_what_2020},~\cite{lella_practical_2022},  ~\cite{dantoni_what_2022},~\cite{abdul_vault_2024},  ~\cite{preet_what_2024},~\cite{shelton_what_2022},  ~\cite{schillerstrom_what_2022},  ~\cite{bairyev_unlocking_2023},~\cite{bremmers_how_2021},  ~\cite{gianchandani_chaos_2022},~\cite{kadikar_building_2023},  ~\cite{bhaskar_chaos_2022},~\cite{eliot_verify_2022},  ~\cite{gunja_what_2024},~\cite{wickramasinghe_chaos_2023},  ~\cite{chattopadhyay_guide_2024},~\cite{hanmer_chaos_2024},  ~\cite{green_7_2023},~\cite{kalal_day50-_2024},  ~\cite{sanwal_chaos_2023},~\cite{treat_guidelines_2020},  ~\cite{ukkuru_chaos_2023},~\cite{keemick_quick_2023},  ~\cite{singh_gill_chaos_2021},~\cite{apexon_what_2020},  ~\cite{tricentis_chaos_2022},~\cite{adservio_chaos_2021},  ~\cite{opentext_what_nodate}
    \\ \hline
    C8      &~\cite{gremlin_chaos_2023},~\cite{andrades_what_2023},  ~\cite{hornsby_what_2020},~\cite{lella_practical_2022},  ~\cite{abdul_vault_2024},~\cite{wakayama_article_2020},  ~\cite{prithvi_chaos_2021},~\cite{bremmers_how_2021},  ~\cite{gianchandani_chaos_2022},~\cite{kadikar_building_2023},  ~\cite{bhaskar_chaos_2022},~\cite{eliot_verify_2022},  ~\cite{gunja_what_2024},~\cite{wickramasinghe_chaos_2023},  ~\cite{green_7_2023},~\cite{kalal_day50-_2024},  ~\cite{sanwal_chaos_2023},~\cite{ukkuru_chaos_2023},  ~\cite{singh_gill_chaos_2021},~\cite{apexon_what_2020},  ~\cite{opentext_what_nodate}
    \\ \hline
    C9      &~\cite{aws_rel12-bp05_nodate},  ~\cite{claytonsiemens77_recommendations_2023},  ~\cite{singh_art_2023},~\cite{andrades_what_2023},  ~\cite{bairyev_unlocking_2023},~\cite{eliot_verify_2022},  ~\cite{wickramasinghe_chaos_2023},~\cite{green_7_2023},  ~\cite{ukkuru_chaos_2023},~\cite{keemick_quick_2023}
    \\ \hline
    C10     &~\cite{aws_rel12-bp05_nodate},  ~\cite{kareliya_what_2023},~\cite{prasla_embracing_2024},  ~\cite{rosenthal_what_2021},~\cite{lella_practical_2022},  ~\cite{wakayama_article_2020},~\cite{schillerstrom_what_2022},  ~\cite{j_haber_why_2021},~\cite{parekh_building_2021},  ~\cite{gianchandani_chaos_2022},~\cite{kadikar_building_2023},  ~\cite{ahamed_chaos_2023},~\cite{eliot_verify_2022},  ~\cite{chattopadhyay_guide_2024},~\cite{green_7_2023},  ~\cite{kalal_day50-_2024},~\cite{sanwal_chaos_2023},  ~\cite{nino_roa_chaos_2022},~\cite{apexon_what_2020},  ~\cite{community_principles_2019}
    \\ \hline
    \end{tabular}
    \vspace{-0.1mm}
\end{table*}

\paragraph{\textbf{Experiments}} This category encompasses the practical execution of CE principles. It includes designing, implementing, and managing experiments to test system resilience. Concepts in this category focus on replicating real-world conditions to assess the system's behavior under stress or failure scenarios. The goal is to uncover weaknesses in the system and improve reliability through iterative testing. This category features five concepts:
\begin{compactenum}
 \item \textbf{C3. Controllably Vary Real-World events.} This concept aims to explain the types of events that can happen in a real system. In the question "What can go wrong?" we can answer in a lot of different ways, but we have to take into consideration the events that can happen in the real environment, not some event that cannot happen in a system. For example, if my system/application is not using a connection, it has no point in considering an event in which the internet halts.
    \item \textbf{C4. Run Experiment in Production.} Basiri et al.~\cite{basiri_chaos_2016} explained this concept as necessary to have real feedback on the system. They also said that only the production environment is the one that can guarantee the most real behavior in case of failure
    \item \textbf{C5. Automate Experiments to Run Continuously.} The purpose of this concept is connected to the fact that the CE experiment should not be run only once but should continue to run, so that in case of differences in the system itself or in the environment, the continuous automated test can help to control without the need for a manual test.
    \item \textbf{C6. Contain the Impact.} This concept is mentioned later in the history of CE. The goal is to contain the impact of experiments conducted on the production system. CE experiments should not disrupt user activities.
    \item \textbf{C8. Increase Complexity Controllably.} This concept is considered important in CE because it relies on starting with basic issues when attempting CE experiments and gradually increasing complexity or involving more parts of the system.
\end{compactenum}

\begin{figure}
    \centering
    \includegraphics[width=1\linewidth]{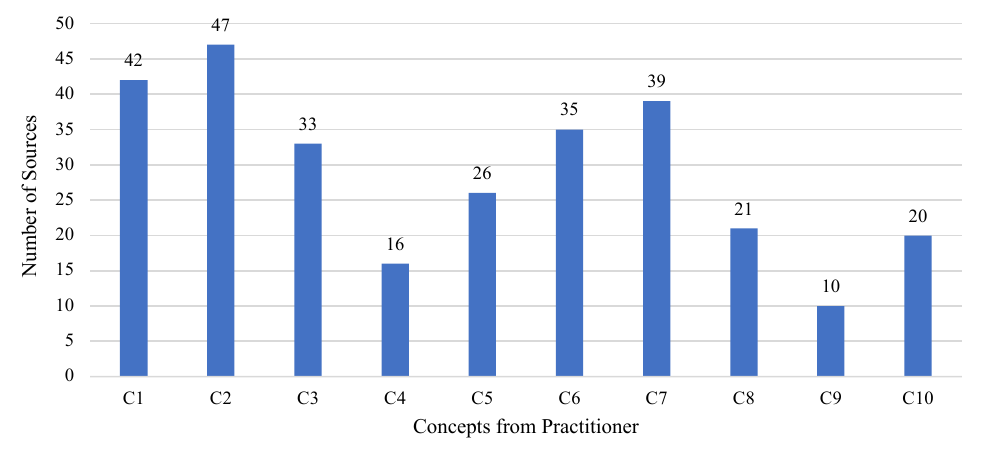}
    \caption{Distribution of the selected gray studies by concepts.}
    \label{fig:practPrinc}
    \vspace{-0.2mm}
\end{figure}

\paragraph{\textbf{Learning and Collaboration}} This category emphasizes the importance of communication, collaboration, and knowledge-sharing within and across teams. CE is not just a technical discipline but a cultural practice that requires engagement from diverse stakeholders. Continuous socialization ensures alignment of goals, facilitates learning, and integrates insights into broader organizational strategies. It highlights the human element of CE, ensuring that experiments drive system and team resilience together. This category features two concepts:
\begin{compactenum}
    \item \textbf{C7. Measure, Learn, Improve.}  This concept aims to include in the classification that CE also aims to measure the system and monitor it in a specific way. CE can lead to greater awareness of the system through a continuous learning process, and the experiments can give outputs that help to improve the system's resilience.
    \item \textbf{C9. Socialize Continuously.}  This concept includes the DevOps environment composed of teams and other components, involving not only the technical but also the human aspect of the system. Therefore, for many, the need to communicate between different teams in an enterprise environment is considered fundamental in CE, while for others, it is not even mentioned.
\end{compactenum}

\subsection{How the elicited concepts are discussed in their sources}
\label{subsec:classification_results}

Table~\ref{tab:concepts_sources} associates the ten concepts to the analyzed resources, whereas Fig.~\ref{fig:practPrinc}  summarizes the distribution of resources for the different concepts. Unsurprisingly, the most frequent concepts are  "C1: Define the steady-state of a system",  "C2: Build a hypothesis", and "C7: Vary real-world events" as tightly related to the key principles of CE. Instead, fewer resources were found for concepts related to "C9: Socialize continuously" and, surprisingly, also for "C4: Run experiments in production", which is, again, a key principle of CE.
In the following, we discuss the main content of the sources that we analyzed. 

\subsubsection{C1. Define the Steady State of the System}
\label{subsubsec:C1}

The sources that did not prioritize it as a main feature---primarily Bremmers~\cite{bremmers_how_2021} and Ukkuru~\cite{ukkuru_chaos_2023}---focus on the experiment procedure, thereby diverting attention away from the definition of the steady state. Also, Chibuike~\cite{chibuike_chaos_2023} emphasizes the importance of having a testing plan while directly addressing hypotheses.
From a different perspective, sources define the baseline or steady state as the system's normal state, characterized by its typical behavior under normal load conditions or when in equilibrium, e.g., Kadikar~\cite{kadikar_building_2023}. This state should be defined using measurable quantities~\cite{preet_what_2024,kareliya_what_2023},~\cite{claytonsiemens77_recommendations_2023}. In particular, Bhaskar~\cite{bhaskar_chaos_2022} specifies that metrics should encompass both business and operational aspects, and reflect on the importance of system observability. System observability is also fundamental~\cite{nino_roa_chaos_2022} since it allows us to know the internal state of the system. 

\begin{resultbox}
\textbf{Finding 1: Chaos-as-a-Test.} DevOps Observability and characterizing normal load conditions are key factors to be made synergic with test planning.
\end{resultbox}

\subsubsection{C2. Formulate a Hypothesis Stemming from the Steady State}
\label{subsubsec:C2}

This is one of the first principles by Basiri et al.~\cite{basiri_chaos_2016} defining that "when we formulate a hypothesis for a chaos-engineering experiment, the hypothesis is about a particular kind of metric".
The definition of the hypothesis is based on what happens or what could happen as reported in Keemick~\cite{keemick_quick_2023}. In particular, the hypothesis must be correlated to a specific scenario (e.g., see Abdul~\cite{abdul_vault_2024} or Varma~\cite{varma_mastering_2024}) in which specific events such as failures or abnormal scenarios are simulated. The hypothesis is to be formulated on what could go wrong when certain events are simulated on the system, while Kadikar~\cite{kadikar_building_2023} specifies to hypothesize the expected output from the system in a specific event, as specific as possible in fact. For Bairyev~\cite{bairyev_unlocking_2023} as well as Singh~\cite{singh_gill_chaos_2021}, the hypothesis must be made on the baseline system, taking into account the system's knowledge, while Hanmer~\cite{hanmer_chaos_2024} predicate to also consider the system's historical data and failure analysis. Hornsby~\cite{hornsby_what_2020} also suggests making the hypothesis with all the human (or potentially even AI) subjects involved in the system/project. Sanwal~\cite{sanwal_chaos_2023}, specify that hypotheses must be both functional and non-functional in the system and are based on different types of output that could occur in a scenario. 
\begin{resultbox}
\textbf{Finding 2. Space-out Chaos Hypothesis.} Hypotheses need to be related to specific scenarios/conditions, and account for historical information, encompassing all \textbf{the system experts} involved.
\end{resultbox}

\subsubsection{C3. Controllably Vary Real-World Events}
\label{subsubsec:C3}

The sources that focus on this principle examine real-world events that occur regularly within the production system, emphasizing those that happen most frequently and may be more critical to the system. These events often affect specific components of the organization, indirectly impacting users. They should closely simulate the actual conditions of the system, involving multiple aspects of the system and considering the system's knowledge.

The types of events might include both failures (e.g., servers crashing, storage malfunctioning, etc. ) and non-failures (e.g., traffic spike, server overload, etc.)~\cite{ruqayya_chaos_2022,wakayama_article_2020,preet_what_2024}.  Sanwal~\cite{sanwal_chaos_2023} remarks that while real-world events are based on the experience and knowledge of the system, such as infrastructure events and data elements, they should also involve one or more parts/components of the system~\cite{ahamed_chaos_2023,ibm_what_2024}.

Bairyev~\cite{bairyev_unlocking_2023} also mention basing scenarios on real events to create variations so that we can test the system on the unfolding of the real event---i.e., exploring the event and the lineage of domino effects stemming from it---while Adservio technologies~\cite{adservio_chaos_2021} emphasize that realistic events are introduced to assess how the system responds to specific disruptions that could occur in production. Conversely, Singh~\cite{singh_art_2023} suggests that failure events should closely mimic the real conditions of the system, while Apexon~\cite{apexon_what_2020} advocates introducing variables that reflect real-world events.

\begin{resultbox}
\textbf{Finding 3. Observing all system components.} Monitoring events that can be of interest to the system, including hardware/external component events, would contribute to better observing possible failure root causes.
\end{resultbox}   

\begingroup
\hypersetup{
citecolor=white
}
\begin{figure}
    \centering
    \begin{tikzpicture}
    \node[anchor=south west] at (0,0) {\includegraphics[width=0.9\linewidth]{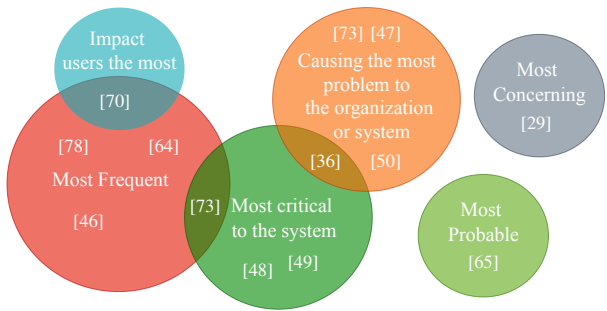}};
    \draw (1.9, 2.9) node[anchor=east] {\color{white}\scriptsize~\cite{bhaskar_chaos_2022}};
    \draw (1.35, 2.3) node[anchor=east] {\color{white}\scriptsize~\cite{ibm_what_2024}};
    \draw (2.55, 2.3) node[anchor=east] {\color{white}\scriptsize~\cite{parekh_building_2021}};
    \draw (1.55, 1.3) node[anchor=east] {\color{white}\scriptsize~\cite{community_principles_2019}};
    \draw (3.1, 1.5) node[anchor=east] {\color{white}\scriptsize~\cite{wickramasinghe_chaos_2023}};
    \draw (3.8, 0.65) node[anchor=east] {\color{white}\scriptsize~\cite{varma_mastering_2024}};
    \draw (4.4, 0.75) node[anchor=east] {\color{white}\scriptsize~\cite{aws_rel12-bp05_nodate}};
    \draw (4.7, 2.1) node[anchor=east] {\color{white}\scriptsize~\cite{chattopadhyay_guide_2024}};
    \draw (5.5, 3.8) node[anchor=east] {\color{white}\scriptsize~\cite{ruqayya_chaos_2022}};
    \draw (5, 3.8) node[anchor=east] {\color{white}\scriptsize~\cite{wickramasinghe_chaos_2023}};
    \draw (5.5, 2.1) node[anchor=east] {\color{white}\scriptsize~\cite{gremlin_chaos_2023}};
    \draw (6.8, 0.75) node[anchor=east] {\color{white}\scriptsize~\cite{prithvi_chaos_2021}};
    \draw (7.5, 2.6) node[anchor=east] {\color{white}\scriptsize~\cite{shelton_what_2022}};
    \end{tikzpicture}
    \caption{Event importance types considered by sources.}
    \label{fig:VaryRealWorld}
    \vspace{-0.1mm}
\end{figure}
\endgroup

   Fig. \ref{fig:VaryRealWorld} illustrates a categorization of event types considered significant by the practitioner of the analyzed resources. The identified categories include the most frequent events, those with the greatest impact on users, the most critical to the system, and those causing the most significant problems for the organization. Additionally, events perceived as the most concerning or the most probable are considered. The size and overlap of the bubbles represent the relationships between these categories and the number of sources that mentioned them.

\subsubsection{C4. Run Experiment in Production}
\label{subsubsec:C4}

Several authors express reservations about conducting chaos experiments directly in production environments due to perceived risks, so they do not consider this concept as a fundamental aspect of CE. Instead, several authors advocate for the utilization of controlled test environments. Eliot et al.~\cite{eliot_verify_2022} describe employing a test environment mirroring the production architecture. In this scenario, the development architecture should closely reflect the production one. This approach proves valuable for companies like Amazon, as it facilitates the development and testing of rollback procedures, ensuring experiments can be stopped when necessary during production testing. Also, Schillerstrom~\cite{schillerstrom_what_2022} argues that testing in production remains the sole method for verifying the true resilience of a system and its architecture, given the presence of real traffic. While Green~\cite{green_7_2023} acknowledges the chaotic nature of production environments, they also suggest conducting chaos experiments in pre-production or isolated (e.g., shadow) environments to mitigate potential risks to users and operations.

Bremmers~\cite{bremmers_how_2021} also contends that CE involves simulating failures in production, whereas J. Haber~\cite{j_haber_why_2021} argues that running experiments in production is the real benefit of CE because the behavior in test environments differs significantly~\cite{hanmer_chaos_2024}. Wickramasinghe~\cite{wickramasinghe_chaos_2023} posit that dev and pre-production systems cannot adequately simulate production due to differences in real loads, while Bhaskar~\cite{bhaskar_chaos_2022} as well as others~\cite{gunja_what_2024,andrades_what_2023} advocate for testing in pre-production and not in production, positing lower risks and maximum gains. Wakayama~\cite{wakayama_article_2020} also recommends conducting chaos experiments in production environments only if the fault injection can be controlled.

OpenText guidelines~\cite{opentext_what_nodate} advise conducting experiments in non-production environments to observe system reactions to specific events, while IBM~\cite{ibm_what_2024} emphasizes the real benefits of running experiments in production environments because it is the environment in which you can obtain the most accurate picture of what can happen with real loads. At the same time, Ruqqaya~\cite{ruqayya_chaos_2022} specifies the importance of conducting experiments in environments that match production as closely as possible, while Singh~\cite{singh_gill_chaos_2021} suggests conducting all experiments in production, but it is possible to test in an environment as close to production as possible.
\begin{resultbox}
\textbf{Finding 4. Production vs. Testing Environment.} Chaos experiments in production are generally preferable if fault injection can be controlled. Otherwise, DevOps may face the challenge of creating as realistic a testing environment as possible.
\end{resultbox}

\subsubsection{C5. Automate Experiments to Run Continuously}
\label{subsubsec:C5}

Apexon~\cite{apexon_what_2020} acknowledges the cyclical nature of the CE process, as evidenced by its discussion of lifecycle management. However, it does not explicitly mention automation as a key concept of CE, despite noting that automation and continuous testing should accompany CE. Similarly, Treat~\cite{treat_guidelines_2020} focuses more on the CE process without explicitly mentioning automation, while D'Antoni~\cite{dantoni_what_2022} recognizes the need for continuous testing but does not highlight automation as a key point. Conversely, Clifford~\cite{clifford_unleash_2022} explicitly states that automation is part of best practices, as do Schillerstrom~\cite{schillerstrom_what_2022} and many others~\cite{wickramasinghe_chaos_2023,kalal_day50-_2024,singh_gill_chaos_2021}, who report it as a principal aspect of CE.

Automation of chaos experiments is viewed as integral to enabling continuous testing within a system. This integration aligns well with DevOps pipelines~\cite{treat_guidelines_2020}, ensuring testing occurs whenever the system is updated. This can be facilitated through specialized tools~\cite{varma_mastering_2024}.

Furthermore, automation is pursued because manual experimentation is labor-intensive, reducing team workload~\cite{arvind_chaos_2021,wickramasinghe_chaos_2023,ibm_what_2024} and consequently lowering costs~\cite{tricentis_chaos_2022} while enabling faster feedback on experiment outputs~\cite{bhaskar_chaos_2022}. Additionally, IBM~\cite{ibm_what_2024} reports that the experiment design, the failure injection, and the infrastructure provisioning can all be automated. The automation allows for regular experimentation, providing up-to-date insights into system resilience~\cite{kalal_day50-_2024} that are able to facilitate the system analysis during orchestration~\cite{community_principles_2019}, and in case of unexpected behavior during the failure simulation, it allows for fixing the problem before it naturally happens in the production environment~\cite{wakayama_article_2020}. Moreover, process automation aids in system improvement and quality enhancement~\cite{abdul_vault_2024,singh_gill_chaos_2021}, and it fosters a culture of CE within the team's responsibilities~\cite{abdul_vault_2024}. Automation also ensures experiment reproducibility and scalability across system components~\cite{green_7_2023}.
\begin{resultbox}
\textbf{Finding 5. Incremental Automation.} While not all sources mention the need for automation, such a need is (unsurprisingly) a key factor for Chaos Engineering. Sources also report that in some circumstances, such an automation may be pursued gradually to minimize risks.
\end{resultbox}

\subsubsection{C6. Contain the Impact}
\label{subsubsec:C6}

Minimizing the so-called blast radius aims to prevent disruptions to normal system operations during CE tests~\cite{andrades_what_2023},~\cite{wakayama_article_2020}, particularly avoiding impacts on users or customers~\cite{tricentis_chaos_2022},~\cite{adservio_chaos_2021}, especially in production environments~\cite{community_principles_2019}. It is crucial not to affect users during testing, especially in production, thus it is advisable to conduct experiments during off-peak hours or when the system can be quickly restored~\cite{wickramasinghe_chaos_2023},~\cite{chattopadhyay_guide_2024},~\cite{ibm_what_2024}.

Hornsby~\cite{hornsby_what_2020} as well as Gunja~\cite{gunja_what_2024} report that the experiment designs can be tailored to limit the blast radius by starting with small experiments~\cite{singh_gill_chaos_2021} so that, in case of unexpected behavior, they do not cause big problems~\cite{gremlin_chaos_2023}. In addition, running them for a limited duration can limit the impact~\cite{j_haber_why_2021}. It is essential to have a precise scope of the blast radius~\cite{schillerstrom_what_2022} and to test on a subset of the system to localize the impact~\cite{bairyev_unlocking_2023},~\cite{ibm_what_2024}. IBM~\cite{ibm_what_2024} and later Kalal~\cite{kalal_day50-_2024} suggest running tests first in pre-production to understand potential impacts before moving to production~\cite{kalal_day50-_2024,community_principles_2019} while also reporting that running in a pre-production environment is a method to control the blast radius.

Many sources also emphasize the importance of having a rollback plan in case experiments impact the system more than anticipated~\cite{hanmer_chaos_2024,community_principles_2019}. For example, reducing the blast radius manually---e.g., diagnosing and shutting down affected containers to impede failure diffusion---helps mitigate short-term negative effects on the system~\cite{prithvi_chaos_2021}.
\begin{resultbox}
\textbf{Finding 6. Blast-Radius Fire-fighting.} Chaos pipelines require DevOps augmentations to limit, wherever possible, the extent of the affliction of its own operations either in pre- or production environments. Whether through manual ``knobs" or special automation, these hooks would be needed to artificially combat the blast-radius effect diffusion.
\end{resultbox}

\subsubsection{C7. Measure, Learn, Improve}
\label{subsubsec:C7}

Varma~\cite{varma_mastering_2024} emphasizes that observability and monitoring of systems through metrics should serve to analyze the system's behavior during experiments and draw conclusions to continuously improve the system's vulnerabilities. AWS~\cite{aws_rel12-bp05_nodate} and others~\cite{singh_gill_chaos_2021,gianchandani_chaos_2022} focus on measuring the system before and after the experiment to validate or refute hypotheses, while Bairyev~\cite{bairyev_unlocking_2023} reports that if something goes wrong, the team should be able to understand the cause thanks to precise metrics and measurements. Measuring the system during experiments is a way to learn more about the system and its behavior~\cite{hanmer_chaos_2024}, but so is their explanation~\cite{andrades_what_2023,shelton_what_2022,hornsby_what_2020}. Furthermore, Preet~\cite{preet_what_2024} highlights how analyzing metrics and observing system behaviors increases the team's confidence in the system. For many others~\cite{wickramasinghe_chaos_2023,chattopadhyay_guide_2024,green_7_2023,keemick_quick_2023}, conducting measured experiments and analyzing the results is a continuous learning process, with the lessons learned requiring extra efforts to become a source of experience for future experiments in the same community. Finally, Kadikar~\cite{kadikar_building_2023} suggests making detailed reports of the system's behavior, including the relevant metrics, to create institutional knowledge.
\begin{resultbox}
\textbf{Finding 7. Community-level chaos institutionalization.} Since chaos engineering is also designed to test the organizational structure, the effective application of chaos engineering to improve process/people metrics in DevOps pipelines requires continuous, community-learning efforts, suggesting also that the institutionalization of good practices—for example, successful use of chaos experiments for specific targets—and knowledge-sharing would play a key role.
\end{resultbox}

\subsubsection{C8. Increase Complexity Controllably}
\label{subsubsec:C8}

When conducting chaos experiments, it is advisable to increase the complexity gradually. The Gremlin team~\cite{gremlin_chaos_2023} suggests starting with smaller experiments, while others~\cite{lella_practical_2022,kadikar_building_2023,wickramasinghe_chaos_2023} recommend increasing them once confidence is gained, managing them with the system until approaching real scenarios. Abdul~\cite{abdul_vault_2024} and Ahamed~\cite{ahamed_chaos_2023} remark that starting with a simple hypothesis that affects only a part of the system and gradually expanding the scope to impact more parts leads to creating more and more complex tests involving multiple parameters.

Andrades~\cite{andrades_what_2023} also recommends increasing complexity gradually to understand the impact of injected chaos into the system, even gradually transitioning from non-production to production during testing. Wakayama~\cite{wakayama_article_2020}  augmented the scope of experiments that involved teams to gain confidence in the system. Prithvi~\cite{prithvi_chaos_2021} also suggests increasing the blast radius of experiments until a satisfactory system representation is achieved, aiding in simulating real-life events. For example, Gunja~\cite{gunja_what_2024} suggests introducing one variable at a time to control the test. After conducting experiments and verifying that everything is okay, Apexon~\cite{apexon_what_2020} suggests increasing the scope and recommends scaling the experiment if no issues are found.
\begin{resultbox}
\textbf{Finding 8. Multivariate complexity control.} Unsurprisingly, the sources found almost unanimously suggest gradually introducing more complex failure scenarios and expanding the scope, e.g., one variable at a time,  as confidence in chaos experiments and system resilience grows.
\end{resultbox}

\subsubsection{C9. Socialize Continuously}
\label{subsubsec:C9}

Many sources focus on technical aspects only, ignoring the organizational ones. Singh~\cite{singh_art_2023} specifies in their analysis that in CE experiments, multiple teams should be involved to have a collaborative and cross-functional approach to the system, a suggestion also echoed by Andrades~\cite{andrades_what_2023} in best practices. However, Green~\cite{green_7_2023} asserts that CE is not just a new technique but also about integrating resilience into the corporate culture. Therefore, when conducting chaos experiments, multiple parts should be involved to execute them effectively and share the results. Also, AWS~\cite{aws_rel12-bp05_nodate} and Andrades~\cite{andrades_what_2023} specify that when experiments are conducted regularly through game days, it demonstrates teams' response to unexpected events. Gunja~\cite{gunja_what_2024} also recognizes the need for a degree of collaboration speed between multiple teams to resolve incidents in production, but does not highlight this as a key point of CE. In this respect, Siemens~\cite{claytonsiemens77_recommendations_2023} suggests working closely with development teams to conduct targeted experiments, measuring also the communication structure parameters at play.
\begin{resultbox}
\textbf{Finding 9.  Community-Level knowledge sharing.} Chaos engineering requires multiple sources of knowledge and experience from the project. This requires a synergic collaboration and knowledge sharing of multiple (e.g., cross-functional) teams.
\end{resultbox}

\subsubsection{C10. Validate Hypothesis}
\label{subsubsec:C10}

After conducting a CE experiment, it is essential, as specified primarily by Prasla and Ma~\cite{prasla_embracing_2024}, to monitor and validate the results against the expected behavior to improve the system's robustness in case differences are observed. Analyzing the results and validating whether the system exhibited unexpected behaviors is a crucial step, as highlighted by Lella~\cite{lella_practical_2022}. Schillerstrom~\cite{schillerstrom_what_2022} underscore that altering the hypothesis of the steady state during an experiment indicates instability, as it is assumed that the system should always operate according to normal behavior, a sentiment echoed by Ahamed~\cite{ahamed_chaos_2023}, who believes that if there is a difference between control and experimental groups, the hypothesis of stable behavior is contradicted. 
\begin{resultbox}
\textbf{Finding 10. Hypothesis fixed beforehand.} When conducting chaos experiments, the set of hypotheses concerning the steady state must be fixed beforehand—and agreed upon, e.g., via before/after scenarios within the team—rather than continuously changed, as the latter would intrinsically indicate instability. 
\end{resultbox}

\section{Industrial Validation}
\label{sec:glr_discussion}

\subsection{Validation Approach}\label{sec:induval}

To understand the practical value of the proposed findings and elicit valuable observations, lessons learned for practitioners, we showcased the gray literature concepts extracted to practitioners that are working on the reliability of the HPC infrastructure of NXP Semiconductors, that are working in a DevOps setting, with a particular interest in the data pipeline that generates real-time data used by the business.  In the competitive and complex semiconductor domain, where the core business is the design of purpose-built, rigorously tested technologies, the reliability of HPC infrastructure is mission-critical and, in turn, this is consistent with the basic assumptions of chaos engineering. The workshop followed the structure of a Delphi study~\cite{dalkey1963experimental} conducted in an initial preparatory session, followed by focus sessions~\cite{kontio2004using} with groups of two people each.  In each session, we presented the results of the gray literature review with the concept extracted. Afterward, a Q\&A session asking for their perception, as well as concrete experience examples, about the concepts we illustrated during the workshop. Finally, two authors analyzed the workshop outcomes and formulated the findings by overlapping them with our gray literature review findings.
In the following, we present the results of the validation. The rest of this section discusses the outcome of this validation in terms of the themes emerging during the workshop and their mapping to standard DevOps practices at NXP Semiconductors.

\subsection{Mapping CE concepts onto DevOps Principles/Practices}
\label{subsec:mapping}

In the following, we outline how the elicited CE concepts map into a practical DevOps environment. A summary of such a mapping is also reported in Table \ref{tab:mapping}.

From a \textit{process} perspective, DevOps aligns business goals with minimized risks and costs while enhancing product quality through frequent, reliable releases. It adopts Agile principles, short feedback cycles, and continuous improvement, replacing traditional hierarchical approvals with lean methods. From a \textit{people} standpoint, DevOps fosters collaboration between developers and operators, promoting knowledge sharing and cross-functional teams that enhance autonomy, collaboration, and communication. The \textit{delivery} perspective emphasizes Continuous Delivery and Deployment, supported by automation, testing, pipelines, and diverse tools, enabling the shift to microservices by a frequent and reliable release process. Finally, from a \textit{run-time} viewpoint, DevOps ensures performance, scalability, resilience, and reliability through continuous monitoring and alerting, ultimately improving security and stability~\cite{leite_survey_2019}.

We mapped the concepts that we have extracted during the classification of the DevOps concepts as summarized in Table \ref{tab:mapping}. Additionally, we identify capabilities from DevOps practices that can be integrated with CE~\cite{senapathi_devops_2018, amaro_capabilities_2023}, along with relevant examples. In addition, the mapping key concepts of CE can be categorized and mapped to specific aspects of the DevOps framework. 

The first CE concept, \textit{Define the Steady State of the System} (C1), falls under the \textbf{\textit{runtime}} perspective. This involves identifying the system’s normal behavior using KPIs obtained through monitoring. Such metrics enable the establishment of a baseline against which deviations can be measured. Amaro et al.~\cite{amaro_capabilities_2023} identify proactive monitoring and observability as key DevOps capabilities.  Monitoring is facilitated by dashboards that visualize system behavior. \textbf{Perceived Industrial Importance:} In an industrial context, the principle is of paramount importance since a DevOps pipeline without its observability would compromise the way in which the runtime is visible against the business metrics and KPIs set forth by the management.

Building on this, the concept of \textit{Build a Hypothesis} (C2) is linked to the \textbf{\textit{process}} category. By proposing "what-if" scenarios, teams can proactively formulate hypotheses that allow for proactive testing and identification of system weaknesses, thereby improving resilience. \textbf{Perceived Industrial Importance:} The "what-if" scenarios, based on the system and domain knowledge, are essential to identify the potential outages that could occur and evaluate the possible system and business behavior in that scenario.

Testing \textit{Real-World Events} (C3), which stems from runtime monitoring and system observations, also aligns with the \textbf{\textit{runtime}} perspective. These events simulate actual conditions based on previously collected data, enabling realistic testing based on the real events that happen to the system. In a DevOps environment, such events can be extracted from postmortem analyses of past incidents~\cite{amaro_capabilities_2023}. \textbf{Perceived Industrial Importance:} In an industrial setting, incidents are systematically collected and analyzed to improve DevOps workflows, reducing mean time to recovery (MTTR) and enhancing operational resilience.

\textit{Experimentation in Production} (C4), another critical concept, evaluates the system’s resilience under operational conditions and is similarly tied to \textbf{\textit{runtime}}. This testing can be enabled with specific tools, such as Chaos Mesh\footnote{https://github.com/chaos-mesh/chaos-mesh} or Gremlin\footnote{https://www.gremlin.com/product/chaos-engineering}. \textbf{Perceived Industrial Importance:} In an industrial context, the production environment is critical, as it directly supports business operations. Controlled experimentation in production is important to validate resilience strategies in real-world conditions while minimizing risk.

\textit{Automation} (C5) relates to the \textbf{\textit{delivery}} perspective. Automated testing within Continuous Integration/Continuous Delivery (CI/CD) pipelines ensures every new software release is rigorously validated, reducing manual effort and increasing reliability. DevOps capabilities encompass automation processes across multiple dimensions, including deployment, testing, monitoring, alerting, and versioning~\cite{amaro_capabilities_2023}. These automation practices can also be applied to CE experiments. 
\textbf{Perceived Industrial Importance:} In a DevOps industrial environment the automation is the key focus to avoid manual effort on recurrent activities that follow patterns, as CE experiments.

\textit{Contain the Impact} (C6) of CE experiments is vital to maintaining normal system functionality and avoiding disruptions for end users. This principle spans both \textbf{\textit{runtime}} and \textbf{\textit{people}} aspects, requiring technical safeguards as well as collaboration among teams. Effective communication and coordination during experiment execution are necessary to ensure that teams are aware of the root causes of unexpected behaviors. Additionally, active monitoring with alert notifications can help stop a CE experiment if the blast radius becomes too large. \textbf{Perceived Industrial Importance:} Containing potential outages in an industrial environment is essential. The larger the blast radius, the greater the impact on business processes and end users.

The principle of \textit{Measure, Learn, Improve} (C7) bridges the \textbf{\textit{process}} and \textbf{\textit{runtime}} perspectives. Measurement focuses on evaluating experiments and system performance, while learning from feedback fosters continuous improvement in processes and resilience strategies. Amaro et al.~\cite{amaro_capabilities_2023} highlight that DevOps adopts a data-driven approach for improvement, supports a learning culture and experimentation, and emphasizes continuous process and workflow refinement.
\textbf{Perceived Industrial Importance:} DevOps aims to enhance both product development and operational processes. A data-driven approach allows organizations to continuously refine their workflows.

\textit{Gradually Increase Complexity} (C8) allows for incremental assessments of system resilience. This approach is closely tied to \textbf{\textit{process}}, with continuous monitoring ensuring the effects are understood in real-time. The modularity typical of DevOps development can also be applied to CE tests, starting with a small set of components before expanding to larger system areas.
\textbf{Perceived Industrial Importance:}
The expansion of the complexity is essential to reach the real complexity of an industrial environment.

\textit{Socialize Continuosly} (C9) emphasizes collaboration, a hallmark of the \textbf{\textit{people}} perspective in DevOps. Effective teamwork and communication are essential for conducting CE experiments and sharing their outcomes across teams. This aligns with the cross-team collaboration and communication capability highlighted by Amaro et al.~\cite{amaro_capabilities_2023}, as seen in retrospective meetings commonly held in DevOps practices. 
\textbf{Perceived Industrial Importance:} In a DevOps industrial environment, teams are interdependent rather than isolated. Effective communication is crucial to preventing bottlenecks caused by cross-team dependencies and ensuring smooth collaboration.

Finally, \textit{Validate the Hypothesis} (C10) falls within the \textbf{\textit{process}} domain. It focuses on using experiment-driven feedback to confirm or challenge assumptions, driving improvements in system quality and robustness. Together, these concepts integrate seamlessly into the DevOps ones, supporting its goals of enhanced reliability and streamlined processes. Comparing the hypothesis with experiment outcomes enables data-driven business decisions, ensuring that resilience improvements are based on empirical evidence rather than assumptions. \textbf{Perceived Industrial Importance:} Industrial systems are inherently complex, making it challenging to fully understand their behavior. Hypothesis validation ensures that system knowledge is tested empirically, enabling data-driven decision-making for improving resilience.

\section{Threats to validity}
\label{sec:threats_of_validity}

\textbf{Construct validity.} The classification of CE principles from gray literature posed a risk of misinterpretation or oversimplification, particularly due to the non-academic nature of these sources. To mitigate this risk, we adhered to systematic gray literature review guidelines, ensuring consistency in how concepts were defined and categorized. 
Furthermore, comparisons with established academic principles of CE were made to validate and align the extracted concepts with recognized definitions, with the help of specialized practitioners at various increments during the final stages of the coding exercise. However, we still acknowledge that what is reported by practitioners may still be based on a \emph{perceived} alignment of our concepts with their practice, rather than actual alignment.

\textbf{Internal Validity.} One of the main risks is the potential omission of important resources during the search process. As previously discussed, in order to mitigate the risk, we formulated a research query based on the research question to address this risk. Additionally, we employed multiple search engines to mitigate the potential bias introduced by relying on a single search engine. To further ensure the reliability of the selected sources, we applied strict inclusion and exclusion criteria to identify the most relevant materials within the gray literature. Finally, the iterative refinement process during data extraction and classification involved multiple reviewers, reducing the impact of subjective interpretations and improving the robustness of the findings.

\textbf{External Validity.} A limitation arises from the fact that gray literature sources predominantly reflect the perspectives of specific industries, such as technology vendors and cloud services, which may reduce the applicability of findings to other sectors or academic contexts. To address this threat, we systematically selected a diverse set of sources spanning various practitioners, domains, and industries. Additionally, we prioritized insights from widely recognized and broadly adopted practices to identify elements with potential industrial relevance, aided by the industrial validation we provided, which is not meant to generalize beyond the considered domain. 

\section{Conclusions}
\label{sec:glr_conclusions}
In this article, we conducted a gray literature review on the practical concepts of Chaos Engineering (CE). We followed the researchers' guidelines in order to do this particular type of literature review using gray sources. We selected 50 gray sources and extracted the data in order to get an overview of the main concept that each practitioner considers about CE; subsequently, we validated our findings in industrial workshops held in conjunction with NXP Semiconductors. 

Practitioners in our validation report that fewer references are made to the need for environmental communication when practicing CE because many efforts focus on technical systems and overlook the organization as a system.  Consequently, the practice of gamedays, where real failures are simulated and teams are tasked with addressing them as if they were real, is gaining traction, aimed at improving communication within teams. We answer our research question of the GLR \textbf{RQ}: \emph{"What generalizable concepts exist to drive chaos engineering?"}: We can say that the concepts of practitioners highlighted most by the analyzed sources go beyond those defined by the discipline of CE. Particularly, the first principle of Basiri~\cite{basiri_chaos_2016}, which we divided into our first two, can be confirmed, as well as the principle of varying real-world events. Contrary to what Basiri~\cite{basiri_chaos_2016} suggests, execution in production is not universally seen as a practical technique, and instead, automation to run continuously has a fair importance. 

\section*{Acknowledgment}
The authors thank NXP Semiconductors for their collaboration. Moreover, Stefano Fossati acknowledges with gratitude the guidance and support provided by Prof. Willem-Jan Van Den Heuvel from TU/e-TiU. Massimiliano Di Penta is supported by the Italian ‘‘PRIN 2022’’ project TRex-SE: ‘‘Trustworthy Recommenders for Software Engineers’’, grant n. 2022LKJWHC.

\balance
\bibliographystyle{IEEEtran}
\bibliography{bibdata} 

% Generated by IEEEtran.bst, version: 1.14 (2015/08/26)
\begin{thebibliography}{10}
\providecommand{\url}[1]{#1}
\csname url@samestyle\endcsname
\providecommand{\newblock}{\relax}
\providecommand{\bibinfo}[2]{#2}
\providecommand{\BIBentrySTDinterwordspacing}{\spaceskip=0pt\relax}
\providecommand{\BIBentryALTinterwordstretchfactor}{4}
\providecommand{\BIBentryALTinterwordspacing}{\spaceskip=\fontdimen2\font plus
\BIBentryALTinterwordstretchfactor\fontdimen3\font minus \fontdimen4\font\relax}
\providecommand{\BIBforeignlanguage}[2]{{%
\expandafter\ifx\csname l@#1\endcsname\relax
\typeout{** WARNING: IEEEtran.bst: No hyphenation pattern has been}%
\typeout{** loaded for the language `#1'. Using the pattern for}%
\typeout{** the default language instead.}%
\else
\language=\csname l@#1\endcsname
\fi
#2}}
\providecommand{\BIBdecl}{\relax}
\BIBdecl

\bibitem{basiri_chaos_2016}
\BIBentryALTinterwordspacing
A.~Basiri, N.~Behnam, R.~de~Rooij, L.~Hochstein, L.~Kosewski, J.~Reynolds, and C.~Rosenthal, ``Chaos {Engineering},'' \emph{IEEE Software}, vol.~33, no.~3, pp. 35--41, May 2016. [Online]. Available: \url{https://ieeexplore.ieee.org/document/7436642}
\BIBentrySTDinterwordspacing

\bibitem{gremlin_state_2021}
\BIBentryALTinterwordspacing
Gremlin, ``\BIBforeignlanguage{en}{State of {Chaos} {Engineering} 2021},'' 2021. [Online]. Available: \url{https://www.gremlin.com/state-of-chaos-engineering/2021}
\BIBentrySTDinterwordspacing

\bibitem{rosenthal_chaos_2020}
C.~Rosenthal and N.~Jones, \emph{Chaos engineering: system resiliency in practice}, first edition~ed.\hskip 1em plus 0.5em minus 0.4em\relax Beijing [China] ; Sebastopol, CA: O'Reilly Media, Inc, 2020.

\bibitem{sommerville_large-scale_2012}
\BIBentryALTinterwordspacing
I.~Sommerville, D.~Cliff, R.~Calinescu, J.~Keen, T.~Kelly, M.~Kwiatkowska, J.~Mcdermid, and R.~Paige, ``\BIBforeignlanguage{en}{Large-scale complex {IT} systems},'' \emph{\BIBforeignlanguage{en}{Communications of the ACM}}, vol.~55, no.~7, pp. 71--77, Jul. 2012. [Online]. Available: \url{https://dl.acm.org/doi/10.1145/2209249.2209268}
\BIBentrySTDinterwordspacing

\bibitem{mulesoft_2024_nodate}
\BIBentryALTinterwordspacing
Mulesoft and Deloitte, ``\BIBforeignlanguage{en}{2024 {Connectivity} {Benchmark} {Report}}.'' [Online]. Available: \url{https://www.mulesoft.com/lp/reports/connectivity-benchmark}
\BIBentrySTDinterwordspacing

\bibitem{statista_global_nodate}
\BIBentryALTinterwordspacing
statista, ``\BIBforeignlanguage{en}{Global hourly enterprise server downtime cost 2019}.'' [Online]. Available: \url{https://www.statista.com/statistics/753938/worldwide-enterprise-server-hourly-downtime-cost/}
\BIBentrySTDinterwordspacing

\bibitem{9831186}
N.~Narayanan, Z.~Chen, B.~Fang, G.~Li, K.~Pattabiraman, and N.~DeBardeleben, ``Fault injection for tensorflow applications,'' \emph{IEEE Transactions on Dependable and Secure Computing}, vol.~20, no.~4, pp. 2677--2695, 2023.

\bibitem{10428037}
H.~Chen, P.~Chen, G.~Yu, X.~Li, and Z.~He, ``Microfi: Non-intrusive and prioritized request-level fault injection for microservice applications,'' \emph{IEEE Transactions on Dependable and Secure Computing}, vol.~21, no.~5, pp. 4921--4938, 2024.

\bibitem{garousi_guidelines_2019}
\BIBentryALTinterwordspacing
V.~Garousi, M.~Felderer, and M.~V. Mäntylä, ``\BIBforeignlanguage{en}{Guidelines for including grey literature and conducting multivocal literature reviews in software engineering},'' \emph{\BIBforeignlanguage{en}{Information and Software Technology}}, vol. 106, pp. 101--121, Feb. 2019. [Online]. Available: \url{https://linkinghub.elsevier.com/retrieve/pii/S0950584918301939}
\BIBentrySTDinterwordspacing

\bibitem{tucker_business_2018}
H.~Tucker, L.~Hochstein, N.~Jones, A.~Basiri, and C.~Rosenthal, ``\BIBforeignlanguage{English}{The {Business} {Case} for {Chaos} {Engineering}},'' \emph{\BIBforeignlanguage{English}{IEEE Cloud Computing}}, vol.~5, no.~3, pp. 45--54, 2018.

\bibitem{simonsson_observability_2021}
J.~Simonsson, L.~Zhang, B.~Morin, B.~Baudry, and M.~Monperrus, ``\BIBforeignlanguage{English}{Observability and chaos engineering on system calls for containerized applications in {Docker}},'' \emph{\BIBforeignlanguage{English}{Future Generation Computer Systems}}, vol. 122, pp. 117--129, 2021.

\bibitem{zhang_automatic_2021}
\BIBentryALTinterwordspacing
L.~Zhang, D.~Tiwari, B.~Morin, B.~Baudry, and M.~Monperrus, ``Automatic {Observability} for {Dockerized} {Java} {Applications},'' Jul. 2021, arXiv:1912.06914 [cs]. [Online]. Available: \url{http://arxiv.org/abs/1912.06914}
\BIBentrySTDinterwordspacing

\bibitem{zhang_maximizing_2022}
L.~Zhang, B.~Morin, B.~Baudry, and M.~Monperrus, ``\BIBforeignlanguage{English}{Maximizing {Error} {Injection} {Realism} for {Chaos} {Engineering} {With} {System} {Calls}},'' \emph{\BIBforeignlanguage{English}{IEEE Transactions on Dependable and Secure Computing}}, vol.~19, no.~4, pp. 2695--2708, 2022.

\bibitem{kesim_identifying_2020}
D.~Kesim, A.~Van~Hoorn, S.~Frank, and M.~Haussler, ``\BIBforeignlanguage{English}{Identifying and prioritizing chaos experiments by using established risk analysis techniques},'' vol. 2020-October, 2020, pp. 229--240, iSSN: 1071-9458.

\bibitem{cotroneo_thorfi_2022}
\BIBentryALTinterwordspacing
D.~Cotroneo, L.~De~Simone, and R.~Natella, ``\BIBforeignlanguage{en}{{ThorFI}: a {Novel} {Approach} for {Network} {Fault} {Injection} as a {Service}},'' \emph{\BIBforeignlanguage{en}{Journal of Network and Computer Applications}}, vol. 201, p. 103334, May 2022. [Online]. Available: \url{https://linkinghub.elsevier.com/retrieve/pii/S1084804522000030}
\BIBentrySTDinterwordspacing

\bibitem{ikeuchi_coverage_2023}
H.~Ikeuchi, A.~Watanabe, and Y.~Takahashi, ``\BIBforeignlanguage{English}{Coverage {Based} {Failure} {Injection} {Toward} {Efficient} {Chaos} {Engineering}},'' vol. 2023-May, 2023, pp. 4571--4577, iSSN: 1550-3607.

\bibitem{frank_misim_2022}
\BIBentryALTinterwordspacing
S.~Frank, L.~Wagner, A.~Hakamian, M.~Straesser, and A.~Van~Hoorn, ``{MiSim}: {A} {Simulator} for {Resilience} {Assessment} of {Microservice}-{Based} {Architectures},'' in \emph{2022 {IEEE} 22nd {International} {Conference} on {Software} {Quality}, {Reliability} and {Security} ({QRS})}.\hskip 1em plus 0.5em minus 0.4em\relax Guangzhou, China: IEEE, Dec. 2022, pp. 1014--1025. [Online]. Available: \url{https://ieeexplore.ieee.org/document/10062468/}
\BIBentrySTDinterwordspacing

\bibitem{poltronieri_chaostwin_2021}
F.~Poltronieri, M.~Tortonesi, and C.~Stefanelli, ``\BIBforeignlanguage{English}{{ChaosTwin}: {A} {Chaos} {Engineering} and {Digital} {Twin} {Approach} for the {Design} of {Resilient} {IT} {Services}},'' 2021, pp. 234--238.

\bibitem{fogli_chaos_2024}
M.~Fogli, C.~Giannelli, F.~Poltronieri, C.~Stefanelli, and M.~Tortonesi, ``\BIBforeignlanguage{English}{Chaos {Engineering} for {Resilience} {Assessment} of {Digital} {Twins}},'' \emph{\BIBforeignlanguage{English}{IEEE Transactions on Industrial Informatics}}, vol.~20, no.~2, pp. 1134--1143, 2024.

\bibitem{alvaro_automating_2016}
\BIBentryALTinterwordspacing
P.~Alvaro, K.~Andrus, C.~Sanden, C.~Rosenthal, A.~Basiri, and L.~Hochstein, ``Automating {Failure} {Testing} {Research} at {Internet} {Scale},'' in \emph{Proceedings of the {Seventh} {ACM} {Symposium} on {Cloud} {Computing}}, ser. {SoCC} '16.\hskip 1em plus 0.5em minus 0.4em\relax New York, NY, USA: Association for Computing Machinery, 2016, pp. 17--28. [Online]. Available: \url{https://dl.acm.org/doi/10.1145/2987550.2987555}
\BIBentrySTDinterwordspacing

\bibitem{basiri_automating_2019}
A.~Basiri, L.~Hochstein, N.~Jones, and H.~Tucker, ``\BIBforeignlanguage{English}{Automating {Chaos} {Experiments} in {Production}},'' 2019, pp. 31--40.

\bibitem{jernberg_getting_2020}
\BIBentryALTinterwordspacing
H.~Jernberg, P.~Runeson, and E.~Engström, ``\BIBforeignlanguage{en}{Getting {Started} with {Chaos} {Engineering} - design of an implementation framework in practice},'' in \emph{\BIBforeignlanguage{en}{Proceedings of the 14th {ACM} / {IEEE} {International} {Symposium} on {Empirical} {Software} {Engineering} and {Measurement} ({ESEM})}}.\hskip 1em plus 0.5em minus 0.4em\relax Bari Italy: ACM, Oct. 2020, pp. 1--10. [Online]. Available: \url{https://dl.acm.org/doi/10.1145/3382494.3421464}
\BIBentrySTDinterwordspacing

\bibitem{garousi_benefitting_2020}
\BIBentryALTinterwordspacing
V.~Garousi, M.~Felderer, M.~V. Mäntylä, and A.~Rainer, ``\BIBforeignlanguage{en}{Benefitting from the {Grey} {Literature} in {Software} {Engineering} {Research}},'' in \emph{\BIBforeignlanguage{en}{Contemporary {Empirical} {Methods} in {Software} {Engineering}}}, M.~Felderer and G.~H. Travassos, Eds.\hskip 1em plus 0.5em minus 0.4em\relax Cham: Springer International Publishing, 2020, pp. 385--413. [Online]. Available: \url{https://doi.org/10.1007/978-3-030-32489-6_14}
\BIBentrySTDinterwordspacing

\bibitem{farace_grey_2010}
\BIBentryALTinterwordspacing
D.~J. Farace and J.~Schöpfel, Eds., \emph{\BIBforeignlanguage{English}{Grey {Literature} in {Library} and {Information} {Studies}}}.\hskip 1em plus 0.5em minus 0.4em\relax K. G. Saur, 2010, accepted: 2017-03-30 23:55. [Online]. Available: \url{https://library.oapen.org/handle/20.500.12657/45661}
\BIBentrySTDinterwordspacing

\bibitem{garousi_need_2016}
\BIBentryALTinterwordspacing
V.~Garousi, M.~Felderer, and M.~V. Mäntylä, ``\BIBforeignlanguage{en}{The need for multivocal literature reviews in software engineering: complementing systematic literature reviews with grey literature},'' in \emph{\BIBforeignlanguage{en}{Proceedings of the 20th {International} {Conference} on {Evaluation} and {Assessment} in {Software} {Engineering}}}.\hskip 1em plus 0.5em minus 0.4em\relax Limerick Ireland: ACM, Jun. 2016, pp. 1--6. [Online]. Available: \url{https://dl.acm.org/doi/10.1145/2915970.2916008}
\BIBentrySTDinterwordspacing

\bibitem{kitchenham_guidelines_2007}
B.~Kitchenham and S.~Charters, ``Guidelines for performing {Systematic} {Literature} {Reviews} in {Software} {Engineering},'' vol.~2, Jan. 2007.

\bibitem{kai_petersen_guidelines_2015}
{Kai Petersen}, K.~Petersen, {Sairam Vakkalanka}, S.~Vakkalanka, {Ludwik Kuźniarz}, and L.~Kuzniarz, ``Guidelines for conducting systematic mapping studies in software engineering : {An} update,'' \emph{Information \& Software Technology}, vol.~64, no.~64, pp. 1--18, Aug. 2015, mAG ID: 1999798506.

\bibitem{kai_petersen_systematic_2008}
{Kai Petersen}, K.~Petersen, {Robert Feldt}, R.~Feldt, {Shahid Mujtaba}, S.~Mujtaba, {Michael Mattsson}, and M.~Mattsson, ``Systematic mapping studies in software engineering,'' \emph{International Conference on Evaluation \& Assessment in Software Engineering}, pp. 68--77, Jun. 2008, mAG ID: 4214443 S2ID: e28bdc373de80d7ec0e64631a89e64fbdcdae230.

\bibitem{shelton_what_2022}
\BIBentryALTinterwordspacing
K.~Shelton. What is chaos engineering? [Online]. Available: \url{https://chaoskyle.com/what-is-chaos-engineering}
\BIBentrySTDinterwordspacing

\bibitem{singh_art_2023}
\BIBentryALTinterwordspacing
S.~Singh. The art of chaos engineering: Unveiling vulnerabilities for stronger systems {\textbar} {LinkedIn}. [Online]. Available: \url{https://www.linkedin.com/pulse/art-chaos-engineering-unveiling-vulnerabilities-stronger-singh/}
\BIBentrySTDinterwordspacing

\bibitem{arvind_chaos_2021}
\BIBentryALTinterwordspacing
P.~Arvind. Chaos engineering. [Online]. Available: \url{https://devopedia.org/chaos-engineering}
\BIBentrySTDinterwordspacing

\bibitem{rosenthal_what_2021}
\BIBentryALTinterwordspacing
C.~Rosenthal. What chaos engineering is (and isn't). [Online]. Available: \url{https://devops.com/what-chaos-engineering-is-and-isnt/}
\BIBentrySTDinterwordspacing

\bibitem{chibuike_chaos_2023}
\BIBentryALTinterwordspacing
N.~Chibuike. Chaos engineering — my take. [Online]. Available: \url{https://aws.plainenglish.io/chaos-engineering-my-take-488603828bca}
\BIBentrySTDinterwordspacing

\bibitem{hornsby_what_2020}
\BIBentryALTinterwordspacing
A.~Hornsby. What is chaos engineering? the art of breaking things purposefully. [Online]. Available: \url{https://www.itnews.com.au/feature/what-is-chaos-engineering-the-art-of-breaking-things-purposefully-555100}
\BIBentrySTDinterwordspacing

\bibitem{preet_what_2024}
\BIBentryALTinterwordspacing
K.~Preet. What is chaos engineering?(examples,pros \& cons). [Online]. Available: \url{https://www.knowledgehut.com/blog/devops/chaos-engineering}
\BIBentrySTDinterwordspacing

\bibitem{chattopadhyay_guide_2024}
\BIBentryALTinterwordspacing
S.~Chattopadhyay. A guide to chaos engineering. [Online]. Available: \url{https://medium.com/@shubhadeepchat/a-guide-to-chaos-engineering-248c220bb9c9}
\BIBentrySTDinterwordspacing

\bibitem{hanmer_chaos_2024}
\BIBentryALTinterwordspacing
S.~Hanmer. Chaos engineering the {AWS} way: 101. [Online]. Available: \url{https://dev.to/aws-builders/chaos-engineering-the-aws-way-101-1acb}
\BIBentrySTDinterwordspacing

\bibitem{green_7_2023}
\BIBentryALTinterwordspacing
S.~Green. (7) {SRE}’s guide to chaos engineering: Embrace the chaos for resilience {\textbar} {LinkedIn}. [Online]. Available: \url{https://www.linkedin.com/pulse/sres-guide-chaos-engineering-embrace-resilience-simon-green/}
\BIBentrySTDinterwordspacing

\bibitem{kalal_day50-_2024}
\BIBentryALTinterwordspacing
S.~Kalal. Day50- chaos engineering: Principles and practices. [Online]. Available: \url{https://sourabhkalal.medium.com/day50-chaos-engineering-principles-and-practices-bf4944becbf5}
\BIBentrySTDinterwordspacing

\bibitem{sanwal_chaos_2023}
\BIBentryALTinterwordspacing
S.~Sanwal. Chaos engineering: Path to build resilient and fault-tolerant software applications - {DZone}. [Online]. Available: \url{https://dzone.com/articles/chaos-engineering-path-to-build-resilient-and-faul}
\BIBentrySTDinterwordspacing

\bibitem{treat_guidelines_2020}
\BIBentryALTinterwordspacing
T.~Treat. Guidelines for chaos engineering, part 1. [Online]. Available: \url{https://blog.realkinetic.com/guidelines-for-chaos-engineering-part-1-e5528a8a219}
\BIBentrySTDinterwordspacing

\bibitem{ukkuru_chaos_2023}
\BIBentryALTinterwordspacing
G.~Ukkuru. Chaos engineering: Principles and best practices for ensuring system resilience. [Online]. Available: \url{https://medium.com/@ukkuru/chaos-engineering-principles-and-best-practices-for-\\ensuring-system-resilience-2a941a0e9e29}
\BIBentrySTDinterwordspacing

\bibitem{keemick_quick_2023}
\BIBentryALTinterwordspacing
Y.~Keemick. A quick guide to chaos engineering {\textbar} by yaël keemink {\textbar} 04.02.2023 {\textbar} to the root. [Online]. Available: \url{https://totheroot.io/article/a-quick-guide-to-chaos-engineering}
\BIBentrySTDinterwordspacing

\bibitem{nino_roa_chaos_2022}
\BIBentryALTinterwordspacing
Y.~Niño~Roa and B.~Linders. Chaos engineering and observability with visual metaphors. [Online]. Available: \url{https://www.infoq.com/articles/chaos-engineering-observability-visual-metaphors/}
\BIBentrySTDinterwordspacing

\bibitem{tricentis_chaos_2022}
\BIBentryALTinterwordspacing
Tricentis. Chaos engineering. [Online]. Available: \url{https://www.tricentis.com/learn/chaos-engineering/}
\BIBentrySTDinterwordspacing

\bibitem{community_principles_2019}
\BIBentryALTinterwordspacing
C.~E. Community. {PRINCIPLES} {OF} {CHAOS} {ENGINEERING} - principles of chaos engineering. [Online]. Available: \url{https://principlesofchaos.org/}
\BIBentrySTDinterwordspacing

\bibitem{ruqayya_chaos_2022}
\BIBentryALTinterwordspacing
N.-u.-A. Ruqayya. Chaos engineering: What is it \& how does it work? [Online]. Available: \url{https://www.blameless.com/blog/chaos-engineering}
\BIBentrySTDinterwordspacing

\bibitem{varma_mastering_2024}
\BIBentryALTinterwordspacing
R.~Varma. Mastering the art of chaos engineering. [Online]. Available: \url{https://www.buildpiper.io/blogs/mastering-the-art-of-chaos-engineering/}
\BIBentrySTDinterwordspacing

\bibitem{aws_rel12-bp05_nodate}
\BIBentryALTinterwordspacing
AWS. {REL}12-{BP}05 test resiliency using chaos engineering - reliability pillar. [Online]. Available: \url{https://docs.aws.amazon.com/wellarchitected/latest/reliability-pillar/rel_testing_resiliency_failure_injection_resiliency.html}
\BIBentrySTDinterwordspacing

\bibitem{gremlin_chaos_2023}
\BIBentryALTinterwordspacing
Chaos engineering: the history, principles, and practice. [Online]. Available: \url{https://www.gremlin.com/community/tutorials/chaos-engineering-the-history-principles-and-practice}
\BIBentrySTDinterwordspacing

\bibitem{claytonsiemens77_recommendations_2023}
\BIBentryALTinterwordspacing
claytonsiemens77. Recommendations for designing a reliability testing strategy - microsoft azure well-architected framework. [Online]. Available: \url{https://learn.microsoft.com/en-us/azure/well-architected/reliability/testing-strategy}
\BIBentrySTDinterwordspacing

\bibitem{kareliya_what_2023}
\BIBentryALTinterwordspacing
A.~K. Kareliya. What is chaos engineering? - its definition, benefits, and best practices. [Online]. Available: \url{https://radixweb.com/blog/what-is-chaos-engineering}
\BIBentrySTDinterwordspacing

\bibitem{prasla_embracing_2024}
\BIBentryALTinterwordspacing
A.~Prasla and J.~Ma. Embracing the chaos: Database resiliency engineering. [Online]. Available: \url{https://www.dell.com/en-us/blog/embracing-the-chaos-database-resiliency-engineering/}
\BIBentrySTDinterwordspacing

\bibitem{maheshwari_chaos_2023}
\BIBentryALTinterwordspacing
A.~Maheshwari. Chaos engineering: How it works? benefits \& challenges. [Online]. Available: \url{https://www.valuelabs.com/resources/blog/software-development/chaos-engineering-guide/}
\BIBentrySTDinterwordspacing

\bibitem{clifford_unleash_2022}
\BIBentryALTinterwordspacing
C.~Clifford. Unleash that chaos (engineering). [Online]. Available: \url{https://rabobank.jobs/en/techblog/unleash-that-chaos-engineering/}
\BIBentrySTDinterwordspacing

\bibitem{andrades_what_2023}
\BIBentryALTinterwordspacing
G.~Andrades. What is chaos engineering? principles, best practices, advantages. [Online]. Available: \url{https://www.accelq.com/blog/chaos-engineering/}
\BIBentrySTDinterwordspacing

\bibitem{lella_practical_2022}
\BIBentryALTinterwordspacing
J.~N. Lella. A practical guide to chaos testing - chaos engineering {\textbar} cigniti. Section: Chaos Engineering. [Online]. Available: \url{https://www.cigniti.com/blog/guide-chaos-engineering/}
\BIBentrySTDinterwordspacing

\bibitem{dantoni_what_2022}
\BIBentryALTinterwordspacing
J.~D'Antoni. What is chaos engineering? history and benefits guide. [Online]. Available: \url{https://orangematter.solarwinds.com/2022/08/18/what-is-chaos-engineering/}
\BIBentrySTDinterwordspacing

\bibitem{abdul_vault_2024}
\BIBentryALTinterwordspacing
K.~Abdul. Vault chaos engineering. [Online]. Available: \url{https://www.hashicorp.com/blog/vault-chaos-engineering}
\BIBentrySTDinterwordspacing

\bibitem{wakayama_article_2020}
\BIBentryALTinterwordspacing
K.~Wakayama. Article detail. [Online]. Available: \url{https://codersociety.com/blog/articles/chaos-engineering}
\BIBentrySTDinterwordspacing

\bibitem{schillerstrom_what_2022}
\BIBentryALTinterwordspacing
M.~Schillerstrom. What is chaos engineering? intro, definition \& more {\textbar} harness. [Online]. Available: \url{https://harness.io/blog/chaos-engineering}
\BIBentrySTDinterwordspacing

\bibitem{bairyev_unlocking_2023}
\BIBentryALTinterwordspacing
M.~Bairyev. Unlocking system reliability and security with chaos engineering. [Online]. Available: \url{https://maddevs.io/blog/chaos-engineering/}
\BIBentrySTDinterwordspacing

\bibitem{j_haber_why_2021}
\BIBentryALTinterwordspacing
M.~J.~Haber. Why now is the time for chaos (engineering) {\textbar} {CSA}. [Online]. Available: \url{https://cloudsecurityalliance.org/blog/2021/11/30/why-now-is-the-time-for-chaos-engineering}
\BIBentrySTDinterwordspacing

\bibitem{parekh_building_2021}
\BIBentryALTinterwordspacing
S.~Parekh and P.~Ramakrishnan. Building resiliency with chaos engineering. [Online]. Available: \url{https://www.thoughtworks.com/en-in/insights/blog/agile-engineering-practices/building-resiliency-chaos-engineering}
\BIBentrySTDinterwordspacing

\bibitem{prithvi_chaos_2021}
\BIBentryALTinterwordspacing
R.~Prithvi. Chaos engineering: Introduction, principles \& evolution. [Online]. Available: \url{https://www.chaosnative.com/blog/defining_chaos_engineering}
\BIBentrySTDinterwordspacing

\bibitem{bremmers_how_2021}
\BIBentryALTinterwordspacing
R.~Bremmers. How implementing chaos engineering can benefit your project. [Online]. Available: \url{https://testdevlab.com/blog/how-implementing-chaos-engineering-can-benefit-your-project}
\BIBentrySTDinterwordspacing

\bibitem{gianchandani_chaos_2022}
\BIBentryALTinterwordspacing
N.~Gianchandani, D.~Anoop~Sahni, and R.~Singh. Chaos engineering best practices to avoid outages. [Online]. Available: \url{https://www.nagarro.com/en/blog/chaos-engineering-best-practices}
\BIBentrySTDinterwordspacing

\bibitem{kadikar_building_2023}
\BIBentryALTinterwordspacing
R.~Kadikar. Building resilience with chaos engineering and litmus. [Online]. Available: \url{https://www.infracloud.io/blogs/building-resilience-chaos-engineering-litmus/}
\BIBentrySTDinterwordspacing

\bibitem{ahamed_chaos_2023}
\BIBentryALTinterwordspacing
I.~Ahamed and S.~Khan. Chaos engineering tutorial: Comprehensive guide with best practices. [Online]. Available: \url{https://www.lambdatest.com/learning-hub/chaos-engineering-tutorial}
\BIBentrySTDinterwordspacing

\bibitem{bhaskar_chaos_2022}
\BIBentryALTinterwordspacing
S.~Bhaskar. Chaos engineering: A step toward reliability. [Online]. Available: \url{https://www.virtusa.com/insights/perspectives/chaos-engineering}
\BIBentrySTDinterwordspacing

\bibitem{eliot_verify_2022}
\BIBentryALTinterwordspacing
S.~Eliot, J.~Barto, and L.~Domb. Verify the resilience of your workloads using chaos engineering {\textbar} {AWS} architecture blog. Section: Architecture. [Online]. Available: \url{https://aws.amazon.com/blogs/architecture/verify-the-resilience-of-your-workloads-using-chaos-engineering/}
\BIBentrySTDinterwordspacing

\bibitem{gunja_what_2024}
\BIBentryALTinterwordspacing
S.~Gunja. What is chaos engineering? [Online]. Available: \url{https://www.dynatrace.com/news/blog/what-is-chaos-engineering/}
\BIBentrySTDinterwordspacing

\bibitem{wickramasinghe_chaos_2023}
\BIBentryALTinterwordspacing
S.~Wickramasinghe. Chaos engineering: Benefits, best practices, and challenges. [Online]. Available: \url{https://www.splunk.com/en_us/blog/learn/chaos-engineering.html}
\BIBentrySTDinterwordspacing

\bibitem{singh_gill_chaos_2021}
\BIBentryALTinterwordspacing
N.~Singh~Gill. Chaos engineering principles ,tools and best practices. [Online]. Available: \url{https://www.xenonstack.com/insights/chaos-engineering}
\BIBentrySTDinterwordspacing

\bibitem{apexon_what_2020}
\BIBentryALTinterwordspacing
Apexon. What is chaos engineering? approaches \& best practices - apexon. [Online]. Available: \url{https://www.apexon.com/resources/white-papers/chaos-engineering/}
\BIBentrySTDinterwordspacing

\bibitem{adservio_chaos_2021}
\BIBentryALTinterwordspacing
Adservio. Chaos engineering best practices. [Online]. Available: \url{https://adservio.fr/post/chaos-engineering-best-practices}
\BIBentrySTDinterwordspacing

\bibitem{opentext_what_nodate}
\BIBentryALTinterwordspacing
OpenText. What is chaos engineering? [Online]. Available: \url{https://www.opentext.com/what-is/chaos-engineering}
\BIBentrySTDinterwordspacing

\bibitem{ibm_what_2024}
\BIBentryALTinterwordspacing
IBM. What is chaos engineering? {\textbar} {IBM}. [Online]. Available: \url{https://www.ibm.com/topics/chaos-engineering}
\BIBentrySTDinterwordspacing

\bibitem{dalkey1963experimental}
N.~Dalkey and O.~Helmer, ``An experimental application of the delphi method to the use of experts,'' \emph{Management science}, vol.~9, no.~3, pp. 458--467, 1963.

\bibitem{kontio2004using}
J.~Kontio, L.~Lehtola, and J.~Bragge, ``Using the focus group method in software engineering: obtaining practitioner and user experiences,'' in \emph{Proceedings. 2004 International Symposium on Empirical Software Engineering, 2004. ISESE'04.}\hskip 1em plus 0.5em minus 0.4em\relax IEEE, 2004, pp. 271--280.

\bibitem{leite_survey_2019}
\BIBentryALTinterwordspacing
L.~Leite, C.~Rocha, F.~Kon, D.~Milojicic, and P.~Meirelles, ``A survey of devops concepts and challenges,'' \emph{ACM Comput. Surv.}, vol.~52, no.~6, Nov. 2019. [Online]. Available: \url{https://doi.org/10.1145/3359981}
\BIBentrySTDinterwordspacing

\bibitem{senapathi_devops_2018}
\BIBentryALTinterwordspacing
M.~Senapathi, J.~Buchan, and H.~Osman, ``\BIBforeignlanguage{en}{{DevOps} {Capabilities}, {Practices}, and {Challenges}: {Insights} from a {Case} {Study}},'' in \emph{\BIBforeignlanguage{en}{Proceedings of the 22nd {International} {Conference} on {Evaluation} and {Assessment} in {Software} {Engineering} 2018}}.\hskip 1em plus 0.5em minus 0.4em\relax Christchurch New Zealand: ACM, Jun. 2018, pp. 57--67. [Online]. Available: \url{https://dl.acm.org/doi/10.1145/3210459.3210465}
\BIBentrySTDinterwordspacing

\bibitem{amaro_capabilities_2023}
\BIBentryALTinterwordspacing
R.~Amaro, R.~Pereira, and M.~M. da~Silva, ``Capabilities and {Practices} in {DevOps}: {A} {Multivocal} {Literature} {Review},'' \emph{IEEE Transactions on Software Engineering}, vol.~49, no.~2, pp. 883--901, Feb. 2023, conference Name: IEEE Transactions on Software Engineering. [Online]. Available: \url{https://ieeexplore.ieee.org/document/9756241/?arnumber=9756241}
\BIBentrySTDinterwordspacing

\end{thebibliography}
\end{document}